\documentclass[12pt,nofootinbib]{revtex4-2}
\usepackage{amsmath}
\usepackage{amsfonts}
\usepackage{amssymb}
\usepackage{amsthm}
 
\usepackage{graphicx}

\begin{document}
\title{The Boltzmann equation and equilibrium thermodynamics
in Lorentz-violating theories}

\author{Robertus Potting$^{1,2}$}

\email{rpotting@ualg.pt}

\affiliation{
$^1$Departamento de F\'\i sica, Faculdade de Ci\^encias e Tecnologia, 
Universidade do Algarve, 8005-139 Faro, Portugal}

\affiliation{
$^2$Centro de Astrof\'\i sica e Gravita\c c\~ao,
Instituto Superior T\'ecnico, Universidade de Lisboa, 
Avenida Rovisco Pais, 1049-001 Lisbon, Portugal}

\begin{abstract}
In this work we adapt the foundations of relativistic kinetic theory and
the Boltzmann equation to particles with Lorentz-violating dispersion relations.
The latter are taken to be those associated to two commonly considered
sets of coefficients in the minimal Standard-Model Extension.
We treat both the cases of classical (Maxwell-Boltzmann) and quantum
(Fermi-Dirac and Bose-Einstein) statistics.
It is shown that with the appropriate definition of the entropy current,
Boltzmann's H-theorem continues to hold.
We derive the equilibrium solutions and then
identify the Lorentz-violating effects for various thermodynamic variables,
as well as for Bose-Einstein condensation.
Finally, a scenario with non-elastic collisions between multiple species of particles
corresponding to chemical or nuclear reactions is considered.
\end{abstract}

\date{\today}

\maketitle

\section{Introduction}

It is a generally accepted notion that the minimal standard model provides
a low-energy limit to a more fundamental theory that includes a quantum
description of gravity. 
Numerous studies have proposed that one of the physical effects of the latter
may be the spontaneous breaking of Lorentz invariance.
A field-theoretic framework that has been used extensively
for studying these is the Standard-Model Extension (SME)
\cite{ColladayKostelecky97,ColladayKostelecky98}.
It contains all possible operators utilizing standard-model fields
that satisfy coordinate reparametrization invariance.
These are generally chosen to preserve energy-momentum conservation,
observer Lorentz invariance, hermiticity, micro-causality and gauge invariance.
Moreover, in its minimal version it is power-counting renormalizable.
These operators are parametrized by coefficients that are now standardized
in the literature.
Based on these coefficients
there have been exhaustive experimental searches for Lorentz and/or CPT violation.
To date these have not encountered any confirmed evidence for Lorentz violation,
and so therefore is reasonable to assume that any violation must be extremely small
in conventional laboratory frames.
For an overview of current bounds, see \cite{KosteleckyRussell}.

Statistical mechanics has been applied in the context of the SME to provide a
mechanism for baryogenesis in thermal equilibrium \cite{Bertolami}.
A general study of statistical mechanics
in the context of the minimal SME to first order on Lorentz-violating coefficients was
performed by Colladay and MacDonald \cite{ColladayMcDonald}.
Other studies of thermodynamics in the presence of SME-type Lorentz violation include 
Refs.\ \cite{deSales:2011jy,Filho:2020cgk,Filho:2020hak,Aguirre:2021tpk}.
We also mention recent studies on the effects of Lorentz violation in the context
of black-hole thermodynamics \cite{Eling:2007qd,Betschart:2008yi,Feldstein:2009qy,Benkel:2018abt}
and of Bose-Einstein condensation \cite{Colladay:2006rt,deSales:2011jy,Casana:2011bv}.

A related topic that, to our knowledge, has been unexplored in the literature
is that of kinetic theory and the Boltzmann equation in the context of Lorentz violation.
In this work we intend to provide a contribution to fill this gap.
To do so we introduce the basic concepts of kinetic theory and
the Boltzmann equation in its relativistic formulations,
but with crucial modifications of the particle dispersion relations due to
Lorentz violation.
The latter are taken to be those associated to two commonly considered
sets of coefficients in the minimal SME,
namely the $a^\mu$ and the $c^{\mu\nu}$
(the former violates CPT, the latter doesn't).
We adopt a observer-Lorentz-invariant setup throughout,
to assure that all constructions are observer independent
and valid to all orders in the Lorentz-violating coefficients.
Both the cases of classical (Maxwell-Boltzmann) and quantum
(Fermi-Dirac and Bose-Einstein) statistics are treated.
We generalize the definition of the entropy current and show that
Boltzmann's H-theorem continues to be valid.
We then derive the equilibrium solutions and from it obtain expressions
for the particle density, the energy density, the isotropic pressure
and the entropy per particle, for fluids with a single species of particle,
and derive the Lorentz-violating effects in Bose-Einstein condensation.
Finally, a scenario is considered with non-elastic collisions between multiple species of particles
corresponding to chemical or nuclear reactions.

While only the cases corresponding to the 
$a^\mu$ and the $c^{\mu\nu}$ coefficients are treated explicitly,
our transparent approach makes a generalization
to other SME-type Lorentz violation straightforward.

This paper is organized as follows.
In section \ref{sec:Boltzmann-equation} we introduce Lorentz-violating dynamics
and derive the relativistic Boltzmann equation,
both for classical (Maxwell-Boltzmann) as well as
quantum (Bose-Einstein and Fermi-Dirac) statistics.
In section \ref{sec:H-theorem} we define the entropy current and
show that Boltzmann's H-theorem applies in the presence of Lorentz violation.
Next, in section \ref{sec:particle-current_EM-tensor},
the particle-number current and the energy momentum tensor are introduced
and their equilibrium properties are evaluated.
In sections \ref{sec:Maxwell-Boltzmann} and \ref{sec:massless} we treat
the special cases of Maxwell-Boltzmann statistics and massless particles,
respectively.
Multiple species of particles are considered in section \ref{sec:multiple}.
Finally, we look at Lorentz-violating effects in Bose-Einstein condensation
in section \ref{sec:Bose-Einstein-condensation},
and then present our conclusions in section \ref{sec:conclusions}.

\section{The Boltzmann equation and Lorentz violation}
\label{sec:Boltzmann-equation}

We consider the relativistic Boltzmann equation for a particle
satisfying a Lorentz-violating dispersion relation
\begin{equation}
D(p^\mu) = \tfrac12 \bigl((\eta^{\mu\lambda}+c^{\mu\lambda}) p_\lambda - a^\mu)
\bigl((\eta_{\mu\kappa}+c_{\mu\kappa})p^\kappa-a_\mu) - \tfrac12 m^2 = 0\>.
\label{dispersion-relation}
\end{equation}
It corresponds to the dispersion relation of a Dirac fermion
in the SME  with nonzero values
of the coefficients $c^{\mu\nu}$ and $a^\mu$.
In this work we will assume that $c^{\mu\nu} = c^{\nu\mu}$.
The dispersion relation \eqref{dispersion-relation} can be derived from 
the covariant particle action
\begin{equation}
S = \int d\tau\, L(\dot x^\mu) =
\int d\tau\left(\tfrac12 e^{-1}\bigl((\eta+c)^{-2}\bigr)_{\mu\nu}\dot{x}^\mu \dot{x}^\nu
+\tfrac12 e m^2 + \dot{x}^\mu \bigl((\eta+c)^{-1}\bigr)_{\mu\nu} A^\nu \right)\>
\label{particle-action}
\end{equation}
where $A_\mu$ is an external field, and we defined $\dot{x}^\mu = dx^\mu/d\tau$. 
The presence of the einbein $e(\tau)$ renders the action invariant
under reparametrizations
\begin{equation}
\tau \to \tau'\>, \qquad e \to e'= \frac{d\tau}{d\tau'}e\>.
\end{equation}
The equation of motion of the einbein yields the constraint
\begin{equation}
\left((\eta+c)^{-2}\right)_{\mu\nu}\dot{x}^\mu \dot{x}^\nu - e^2 m^2 = 0
\label{constraint}
\end{equation}
which can be expressed in terms of the canonical momentum
\begin{equation}
p_\mu = \frac{\partial L}{\partial\dot{x}^\mu} =
e^{-1}\left((\eta+c)^{-2}\right)_{\mu\nu}\dot{x}^\nu + \left((\eta+c)^{-1}\right)_{\mu\nu}a^\nu
\label{pmu_dot-x}
\end{equation}
as Eq.\ \eqref{dispersion-relation},
upon identifying $A_\mu$ with the vector coefficient $a_\mu$.
It is useful to note that the covariant Hamiltonian 
\begin{equation}
H(p^\mu) = p_\mu \dot{x}^\mu - L(\dot{x}^\mu)
\end{equation}
is equal to $e D(p^\mu)$ and therefore vanishes by the einbein equation of motion.

For the velocity we find
\begin{equation}
v^i = \frac{dx^i}{dx^0} = \frac{\dot{x}^i}{\dot{x}^0}=
\frac{(\eta^{i\nu}+c^{i\nu})\bigl((\eta_{\nu\lambda}+c_{\nu\lambda})p^\lambda-a_\nu\bigr)}{(\eta^{0\nu}+c^{0\nu})\bigl((\eta_{\nu\lambda}+c_{\nu\lambda})p^\lambda-a_\nu\bigr)}
= \frac{\tilde{p}^i}{\tilde{p}^0}\>,
\label{vi}
\end{equation}
where we defined the observer four-vector
\begin{equation}
\tilde{p}^\mu = e^{-1} \dot{x}^\mu = (\eta^{\mu\nu}+c^{\mu\nu})\bigl((\eta_{\nu\lambda}+c_{\nu\lambda})p^\lambda-a_\nu\bigr)\>.
\label{tilde-p}
\end{equation}
We will now fix the gauge of reparametrization invariance
by setting the einbein equal to the conventional choice
\begin{equation}
e = \frac{1}{m}
\label{reparametrization-gauge}
\end{equation}
In the absence of Lorentz violation this corresponds to taking $\tau$ equal to the proper time on the particle
(see Eq.\ \eqref{constraint}).
Note that the gauge \eqref{reparametrization-gauge} is only possible for nonzero mass.
In section \ref{sec:massless} we will analyse the massless case.

It then follows from Eq.\ \eqref{pmu_dot-x} that 
\begin{equation}
d\tau = \frac{m dx^0}{\tilde{p}^0}
\label{dtau}
\end{equation}
which is evidently a Lorentz scalar.

We want to consider the Boltzmann equation in the presence of the Lorentz-violating
coefficients $c^{\mu\nu}$ and $a^\mu$,
describing a relativistic gas of particles satisfying the dispersion
relation \eqref{dispersion-relation}.
We will need the relativistic version of the Boltzmann equation;
for an introduction, see, e.g., \cite{CercignaniKremer}.
Basic ingredient is the one-particle phase-space distribution function
$f(\vec{x},\vec{p},t)$, defined such that
\begin{equation}
dN(t) = \frac{g}{h^3} f(\vec{x},\vec{p},t)\,d^3x\,d^3p
\label{dN}
\end{equation}
denotes the number of particles in the volume element $d^3x$ about $\vec{x}$
with momenta in a range $d^3p$ around $\vec{p}$ at time $t$.
Here $h$ is Boltzmann's constant,
while $g$ is the degeneracy factor of the quantum states;
for massive particles with spin $s$, we have $g = 2s+1$.

It is important that we set up a consistent description invariant under
observer Lorentz-transformations.
First of all, note that 
\begin{equation}
d^4p \,\delta\bigl(D(p^\mu)\bigr) =
\frac{d^3p}{\tilde{p}^0}
\label{invariant-3momentum-element}
\end{equation}
is the correct Lorentz-covariant generalization of the usual expression
$d^3p/p^0$.
Noting, furthermore, that $d^4x$ as well as,
from Eq.\ \eqref{dtau},
\begin{equation}
\frac{dx^0}{\tilde{p}^0}
\end{equation}
is Lorentz covariant, it follows that this is also true for
\begin{equation}
d^3x\,\tilde{p}^0\>.
\label{invariant-3space-element}
\end{equation}
Combining Eqs.\ \eqref{invariant-3momentum-element} and \eqref{invariant-3space-element},
it follows that the phase-space element $d^3x\, d^3p \equiv d^6\xi$ is invariant
under observer Lorentz transformations.

From Eq.\ \eqref{dN} it follows
\begin{equation}
dN(t+dt) =
 \frac{g}{h^3} f\left(\vec x + dt\frac{d\vec x}{dt},\vec p + dt\frac{d\vec p}{dt},t+dt\right)
   d^6\xi(t+dt)
\end{equation}
If we assume a force field $\vec{F}(\vec{x},\vec{p},t)$ such that
\begin{equation}
\frac{d\vec{p}}{dt} = \vec{F}
\label{Newton1}
\end{equation}
we have, with Eq.\ \eqref{vi} and
\begin{align}
d^6\xi(t+dt) &= d^6\xi(t) \,\text{det}\left(\frac{(\partial(\vec{x}+\vec v t,\vec{p}+\vec{F} t)}{\partial(\vec{x},\vec{p})}\right) \nonumber\\
&= d^6\xi\left(1 + dt\left(\frac{\partial}{\partial\vec x}\cdot\vec{v} + \frac{\partial}{\partial\vec p}\cdot \vec F\right) + \mathcal{O}(dt^2)\right)
\end{align}
that
\begin{equation}
dN(t+dt)-dN(t)= \frac{g}{h^3} d^6\xi\, dt\left(\frac{\partial f}{\partial t} + 
\frac{\partial}{\partial\vec x}\cdot(\vec v f)
+ \frac{\partial}{\partial\vec p}\cdot(\vec F f)\right)\>.
\label{dN(t+dt)-dN(t)}
\end{equation}
From Eq.\ \eqref{vi} it follows that $\vec{v}$ depends only on $\vec{p}$,
not on $\vec{x}$,
and therefore the first two terms on the right-hand side can be written as
\begin{equation}
\frac{\partial f}{\partial t} + \vec{v}\cdot\frac{\partial f}{\partial\vec x} =
\frac{\partial f}{\partial t} + \frac{\tilde{p}^i}{\tilde{p}^0}
\frac{\partial f}{\partial x^i}
= \frac{\tilde{p}^\mu \partial_\mu f}{\tilde{p}^0}\>.
\end{equation}
Writing Newton's law \eqref{Newton1} in covariant form
\begin{equation}
\frac{dp^\mu }{d\tau} = K^\mu\>.
\end{equation}
it follows, using Eq.\ \eqref{dtau}, that
\begin{equation}
dt \frac{\partial}{\partial \vec{p}}\cdot(\vec{F}f)
= d\tau\frac{\partial(K^\mu f)}{\partial p^\mu}\>.
\end{equation}
Altogether Eq.\ \eqref{dN(t+dt)-dN(t)} is then written as
\begin{equation}
dN(t+dt)-dN(t)= \frac{g}{h^3} d^6\xi\, d\tau\left(
\frac{1}{m}
\tilde{p}^\mu \frac{\partial f}{\partial x^\mu} + 
\frac{\partial(K^\mu f)}{\partial p^\mu}\right)
\label{dN(t+dt)-dN(t)_2}
\end{equation}
from which we find, using again Eq.\ \eqref{dtau}, that
\begin{equation}
\frac{d}{dt}dN = \frac{m}{\tilde{p}^0}
\frac{g}{h^3} d^6\xi
\left(\frac{1}{m}
\tilde{p}^\mu \frac{\partial f}{\partial x^\mu} + 
\frac{\partial(K^\mu f)}{\partial p^\mu}\right)
\label{ddt_dN}
\end{equation}
In the absence of collisions, $dN(t+dt)=dN(t)$ by construction,
and thus the right-hand side of Eq.\ \eqref{ddt_dN} will then vanish.
In the presence of collisions, however,
there will be particles that are scattered out of the volume element,
and others that are scattered in to it, and thus
\begin{equation}
dN = dN^+ - dN^-\>.
\label{dN+-}
\end{equation}
In order to determine its form, we follow Boltzmann's \textit{Sto\ss zahlansatz},
according to which
\begin{itemize}
\item the gas is sufficiently dilute so that only elastic collisions between pairs
of particles are relevant;
\item the momenta of two colliding particles before the collision are uncorrelated
(molecular chaos assumption);
\item the one-particle distribution function varies slowly over a time interval
which is much larger than the duration of the collision but much smaller than the time between
collisions,
and neither does $f$ change much over a distance of the order of the interaction range.
\end{itemize}
The number of collisions $dN_\text{\tiny coll}(p_1,p_2\rightarrow p_1',p_2')$
of particles with momenta around $\vec{p}_1$ and $\vec{p}_2$
scattering to particles with momenta around $\vec{p}'_1$ and $\vec{p}'_2$
should be proportional to $f_1(\vec{x},\vec{p}_1,t) d^3p_1\,f_2(\vec{x},\vec{p}_2,t) d^3p_2$
and also to $d^3x\,dt$ and $d^3p'_1\,d^3p'_2$.
Imposing as well that it should be a Lorentz-covariant expression,
we can express it in terms of the invariant transition rate per unit volume:
\begin{align}
dN_\text{\tiny coll}(p_1,p_2\rightarrow p_1',p_2') &= d^4x \frac{d^3p_1}{\tilde{p}_{01}}
\frac{d^3p_2}{\tilde{p}_{02}}\frac{d^3p'_1}{\tilde{p}'_{01}}
\frac{d^3p'_2}{\tilde{p}'_{02}}\nonumber\\
&\qquad{}\times \frac{g^2}{h^6} f_1(\vec{x},\vec{p}_1,t) f_2(\vec{x},\vec{p}_2,t) W(p_1,p_2\rightarrow p_1',p_2')
\label{dN-coll}
\end{align}
%
%
%
Because of momentum conservation the invariant
$W(p_1,p_2\rightarrow p_1',p_2')$ is proportional to
$\delta^4(p_1' + p_2' - p_1 - p_2)$.
In the absence of Lorentz violation it can be related to the differential scattering
cross section through
\begin{equation}
d\sigma = \frac{d^3p'_1}{p'_{01}}\frac{d^3p'_2}{p'_{02}}
\frac{W(p_1,p_2\rightarrow p_1',p_2')}{v_{\text{\tiny rel}}\,p_1\cdot p_2}\>,
\end{equation}
where the invariant $v_{\text{\tiny rel}}$ is equal to
the relative velocity between particles 1 and 2 in the
``laboratory frame'' where $\vec{p}_2 = 0$ \cite{CercignaniKremer}.
Identifying $p^\mu \equiv p_1^\mu$,
it follows that $dN_1^-$ is equal to expression \eqref{dN-coll} integrated
over all values of $p_2$, $p_1'$ and $p_2'$.
An analogous expression follows for $dN_1^+$,
after switching primed and unprimed momenta.
It then follows that
\begin{align}
\frac{dN_1}{dt} = \frac{dN_1^+}{dt} - \frac{dN_1^-}{dt}
&= d^6\xi_1\frac{1}{\tilde{p}_{01}}
\int\frac{d^3p_2}{\tilde{p}_{02}}\frac{d^3p'_1}{\tilde{p}'_{01}}
\frac{d^3p'_2}{\tilde{p}'_{02}}\nonumber\\
&\quad{}\times\frac{g^2}{h^6} \bigl(f_{1'} f_{2'} W(p_1',p_2'\rightarrow p_1,p_2)
- f_{1} f_{2} W(p_1,p_2\rightarrow p_1',p_2')\bigr)\>.
\label{ddN^+-dN^-}
\end{align}
An important assumption we will now make is that of \textit{detailed balance}:
\begin{equation}
W(p_1',p_2'\rightarrow p_1,p_2) = W(p_1,p_2\rightarrow p_1',p_2')\>.
\label{detailed-balance}
\end{equation}
From Eqs.\ \eqref{ddt_dN}, \eqref{dN+-} and \eqref{ddN^+-dN^-} we then find the \textit{Boltzmann equation}
in covariant form
\begin{equation}
\tilde{p}_1^\mu\frac{\partial f}{\partial x^\mu} + m\frac{\partial(K^\mu f)}{\partial p_1^\mu}
= \frac{g}{h^3} \int\frac{d^3p_2}{\tilde{p}_{02}}\frac{d^3p'_1}{\tilde{p}'_{01}}\frac{d^3p'_2}{\tilde{p}'_{02}}
(f_{1'} f_{2'} - f_{1} f_{2})W(p_1,p_2\rightarrow p_1',p_2')\>.
\label{Boltzmann-eq}
\end{equation}
Eq.\ \eqref{Boltzmann-eq} is valid for a ``classical gas'', that is,
one in which the quantum states have low occupation numbers.

\subsection{The cases of Fermi-Dirac and Bose-Einstein statistics}

Eq.\ \eqref{Boltzmann-eq} can be generalized to a single degenerate fermion gas satisfying the Pauli
exclusion principle by noting that a phase space element $d^6\xi$ is completely occupied
if the number of particles in it equals the number of available states, i.e.,
\begin{equation}
\frac{g}{h^3} f\,d^6\xi = \frac{g}{h^3}d^6\xi \qquad\Rightarrow\qquad f = 1\>.
\end{equation}
Consequently, $1-f$ gives the proportion of vacant states in a phase space element.
Therefore, Eq.\ \eqref{Boltzmann-eq} can be generalized to the case of a gas of fermions by making
the substitution 
\begin{equation}
f_{1'} f_{2'} \to f_{1'} f_{2'}\bigl(1-f_1\bigr)\bigl(1-f_2\bigr)
\end{equation}
in the ``gain'' term, and the substitution
\begin{equation}
f_{1} f_{2} \to f_{1} f_{2}\bigl(1-f_{1'}\bigr)\bigl(1-f_{2'}\bigr)
\end{equation}
in the ``leave'' term.
For bosons it can be shown that the factor $(1-f)$ must be replaced by $(1+f)$
\cite{SnokeLiuGirvin}.
This way we obtain the \textit{relativistic Uehling-Uhlenbeck equation} 
\begin{align}
\tilde{p}_1^\mu\frac{\partial f}{\partial x^\mu} + m\frac{\partial(K^\mu f)}{\partial p_1^\mu}
&= \frac{g}{h^3}\int\frac{d^3p_2}{\tilde{p}_{02}}\frac{d^3p'_1}{\tilde{p}'_{01}}\frac{d^3p'_2}{\tilde{p}'_{02}}
\Biggl[f_{1'} f_{2'}\left(1+\epsilon f_1\right)\left(1+\epsilon f_2\right)\nonumber\\
&\qquad{} - f_{1} f_{2}\left(1+\epsilon f_{1'}\right)\left(1+\epsilon f_{2'}\right)\Biggr]
W(p_1,p_2\rightarrow p_1',p_2')\nonumber\\
&= \frac{g}{h^3}\int\frac{d^3p_2}{\tilde{p}_{02}}\frac{d^3p'_1}{\tilde{p}'_{01}}\frac{d^3p'_2}{\tilde{p}'_{02}}
\bigl(f_{1'} f_{2'}\bar{f}_1\bar{f}_2 - f_1 f_2\bar{f}_{1'}\bar{f}_{2'}\bigr)
W(p_1,p_2\rightarrow p_1',p_2')
\label{UU-eq}
\end{align}
where $\epsilon = +1$ or $-1$, for Bose-Einstein or Fermi-Dirac statistics,
respectively, and
\begin{equation}
\bar{f} = 1 + \epsilon f\>.
\label{bar-f}
\end{equation}
The case of Eq.\ \eqref{Boltzmann-eq} representing Maxwell-Boltzmann statistics
corresponds to the case $\epsilon = 0$.

From Eq.\ \eqref{UU-eq} we obtain the so-called equation of transfer by multiplying
with an arbitrary function $\psi(\vec{x}^\mu,\vec{p}_1^\mu)$ and integrating with respect
to $d^3p_1/\tilde{p}_{01}$. It follows that
\begin{equation}
\int \frac{d^3p_1}{\tilde{p}_{01}} \psi(x,p_1)\left[\tilde{p}_1^\mu\partial f_1
+ m\frac{\partial(K^\mu f)}{\partial p_1^\mu}\right] = \text{Coll}[\psi]
\label{M1}
\end{equation}
with on the right-hand side the collisional functional 
\begin{align}
\text{Coll}[\psi] &= \frac{g}{h^3}\int \frac{d^3p_1}{\tilde{p}_{01}}\frac{d^3p_2}{\tilde{p}_{02}}
\frac{d^3p'_1}{\tilde{p}'_{01}}\frac{d^3p'_2}{\tilde{p}'_{02}} \psi(x,p_1)
W(p_1,p_2\rightarrow p_1',p_2')(f_{1'} f_{2'}\bar{f}_1\bar{f}_2 - f_1 f_2\bar{f}_{1'}\bar{f}_{2'})\nonumber\\
&= \frac{g}{h^3}\int \frac{d^3p_1}{\tilde{p}_{01}}\frac{d^3p_2}{\tilde{p}_{02}}
\frac{d^3p'_1}{\tilde{p}'_{01}}\frac{d^3p'_2}{\tilde{p}'_{02}} (\psi_{1'} - \psi_1)W(p_1,p_2\rightarrow p_1',p_2')f_1 f_2\bar{f}_{1'}\bar{f}_{2'}\nonumber\\
&=\frac{g}{2h^3}\int \frac{d^3p_1}{\tilde{p}_{01}}\frac{d^3p_2}{\tilde{p}_{02}}
\frac{d^3p'_1}{\tilde{p}'_{01}}\frac{d^3p'_2}{\tilde{p}'_{02}} (\psi_{1'} + \psi_{2'} - \psi_1 - \psi_2)
W(p_1,p_2\rightarrow p_1',p_2')f_1 f_2\bar{f}_{1'}\bar{f}_{2'}
\label{Coll-psi}
\end{align}
where in the second equality we used detailed balance \eqref{detailed-balance}.
By rewriting the partial derivative on the left-hand side of Eq.\ \eqref{M1}
with respect to $x^\mu$ in the first term,
and doing a partial integration with respect to $p_1^\mu$ in the second term,
we obtain the \textit{transfer equation}
\begin{equation}
\frac{\partial}{\partial x^\mu}\left( \int \frac{d^3p_1}{\tilde{p}_{01}} f_1\psi_1\tilde{p}_1^\mu\right)
- \int \frac{d^3p_1}{\tilde{p}_{01}}\left(\tilde{p}_1^\mu\frac{\partial\psi_1}{\partial x^\mu} + m K_1^\mu\frac{\partial\psi_1}{\partial p_1^\mu}\right) = \text{Coll}[\psi]\>.
\label{transfer-equation}
\end{equation}

\section{The H-theorem}
\label{sec:H-theorem}

In this section we will derive the so-called H-theorem \cite{Boltzmann}
in the presence of the Lorentz-violating effects.

First we consider the case $\epsilon = 0$.
Take $\psi(\xi) = -\ln\bigl(f(\xi)\bigr) + 1$ in the transfer equation;
it follows that
\begin{align}
&\frac{\partial}{\partial x^\mu}\left( -\int \frac{d^3p_1}{\tilde{p}_{10}} f_1(\ln f_1 -1)\tilde{p}_1^\mu\right)
+ \int \frac{d^3p_1}{\tilde{p}_{10}} \left( \tilde{p}_1^\mu\frac{\partial f_1}{\partial x^\mu} + m K_1^\mu\frac{\partial f_1}{\partial p_1^\mu}\right)\nonumber\\
&\qquad = - \frac{g}{2h^3}\int \frac{d^3p_1}{\tilde{p}_{01}}\frac{d^3p_2}{\tilde{p}_{02}}
\frac{d^3p'_1}{\tilde{p}'_{01}}\frac{d^3p'_2}{\tilde{p}'_{02}} \ln\left(\frac{f_{1'} f_{2'}}{f_1 f_2}\right)
W(p_1,p_2\rightarrow p_1',p_2')f_1 f_2
\label{BMeq1_MB}
\end{align}
From Eq.\ \eqref{M1} we see that the second term on the left-hand side equals Coll[1],
which vanishes identically.
Defining the entropy-density four-current
\begin{equation}
s^\mu(x) = - \frac{kg}{h^3}\int\frac{d^3p}{\tilde{p}_0} \tilde{p}^\mu f(x,p)(\ln f(x,p) - 1)
\label{entropy-density-current_MB}
\end{equation}
where $k$ is Boltzmann's constant,
we find from Eq.\ \eqref{BMeq1_MB}, after adding $\frac{1}{2}\text{Coll}[1] = 0$ to the right-hand side:
\begin{equation}
\frac{\partial s^\mu}{\partial x^\mu} =  \frac{kg}{2h^3}\int\frac{d^3p_1}{\tilde{p}_{01}}\frac{d^3p_2}{\tilde{p}_{02}}
\frac{d^3p'_1}{\tilde{p}'_{01}}\frac{d^3p'_2}{\tilde{p}'_{02}} f_1 f_2
\left[\frac{f_{1'} f_{2'}}{f_1 f_2} - \ln\left(\frac{f_{1'} f_{2'}}{f_1 f_2}\right) - 1\right]
W(p_1,p_2\rightarrow p_1',p_2') \>.
\label{div-S_MB}
\end{equation}

Next consider the cases $\epsilon = \pm 1$.
Under transformation $f \to -\epsilon - f$, or, equivalently
\begin{equation}
f \to -\epsilon \bar{f} \qquad\text{and}\qquad \bar{f} \to -\epsilon f\>,
\end{equation}
with $\bar{f}$ given by Eq.\ \eqref{bar-f},
the collisional functional $\text{Coll}[\psi]$ given by Eq.\ \eqref{Coll-psi} switches sign.
Therefore we have from the transfer equation \eqref{transfer-equation}
\begin{align}
&\frac{\partial}{\partial x^\mu}\left(\int\frac{d^3p_1}{\tilde{p}_1^0}\,f_1\psi\tilde{p}_1^\mu\right) -
\int\frac{d^3p_1}{\tilde{p}_1^0}\,f_1\left(\tilde{p}_1^\mu\frac{\partial\psi}{\partial x^\mu}
+ m K_1^\mu\frac{\partial\psi}{\partial p_1^\mu}\right)\nonumber\\
&\qquad = \epsilon\frac{\partial}{\partial x^\mu}\left(\int\frac{d^3p_1}{\tilde{p}_1^0}\,\bar{f}_1\psi\tilde{p}_1^\mu\right)
- \epsilon\int\frac{d^3p_1}{\tilde{p}_1^0}\,\bar{f}_1\left(\tilde{p}_1^\mu\frac{\partial\psi}{\partial x^\mu}
+ m K_1^\mu\frac{\partial\psi}{\partial p_1^\mu}\right)
\label{anti-symmetry}
\end{align}
for arbitrary function $\psi(x,p_1)$.
Now take $\psi_1 = -\ln f_1$, $\bar{\psi}_1 = -\ln \bar{f}_1$, and consider the expression
\begin{equation}
\frac{\partial}{\partial x^\mu}\left(\int\frac{d^3p_1}{\tilde{p}_1^0}\,f_1(\psi_1 - \bar{\psi}_1)\tilde{p}_1^\mu\right) -
\int\frac{d^3p_1}{\tilde{p}_1^0}\,f_1\left(\tilde{p}_1^\mu\frac{\partial(\psi_1 - \bar{\psi}_1)}{\partial x^\mu}
+ m K_1^\mu\frac{\partial(\psi_1 - \bar{\psi}_1)}{\partial p_1^\mu}\right)\>.
\label{psi-barpsi_expression}
\end{equation}
From the transfer equation \eqref{transfer-equation} it follows that this is equal to Coll$[\psi - \bar{\psi}]$. 
On the other hand,
by applying identity \eqref{anti-symmetry} to the terms involving $\bar{\psi}_1$ 
we see that expression \eqref{psi-barpsi_expression} is equal to
\begin{align}
\frac{\partial}{\partial x^\mu}\left(\int\frac{d^3p_1}{\tilde{p}_1^0}
(f_1\psi_1 - \epsilon\bar{f}_1\bar{\psi}_1)\tilde{p}_1^\mu\right)
& {} -
\int\frac{d^3p_1}{\tilde{p}_1^0}\,f_1\left(\tilde{p}_1^\mu\frac{\partial\psi_1}{\partial x^\mu}
+ m K_1^\mu\frac{\partial\psi_1}{\partial p_1^\mu}\right)\nonumber\\
&{}+ \epsilon\int\frac{d^3p_1}{\tilde{p}_1^0}\,\bar{f}_1\left(\tilde{p}_1^\mu\frac{\partial\bar{\psi}_1}{\partial x^\mu}
+ m K_1^\mu\frac{\partial\bar{\psi}_1}{\partial p_1^\mu}\right)
\end{align}
The second and third terms are both proportional to Coll$[1] = 0$.
Defining the entropy-density four-current
\begin{equation}
s^\mu(x) = \frac{kg}{h^3}\int\frac{d^3p}{\tilde{p}_0} \tilde{p}^\mu 
\left[\epsilon^{-1}\bar{f}\ln\bar{f} - f \ln f \right]
\label{entropy-density-current}
\end{equation}
the first term is proportional to $\partial s^\mu/\partial x^\mu$.
Applying the transfer equation it follows,
upon adding $\frac{1}{2}\text{Coll}[1] = 0$ to the right-hand side, that
\begin{equation}
\frac{\partial s^\mu}{\partial x^\mu} =
\frac{kg}{2h^3}\int\frac{d^3p_1}{\tilde{p}_{01}}\ldots\frac{d^3p'_2}{\tilde{p}'_{02}}f_1 f_2\bar{f}_{1'}\bar{f}_{2'}
\left[\frac{f_{1'} f_{2'}\bar{f}_1\bar{f}_2}{f_1 f_2\bar{f}_1'\bar{f}_2'} - \ln\left(\frac{f_{1'} f_{2'}\bar{f}_1\bar{f}_2}{f_1 f_2\bar{f}_{1'}\bar{f}_{2'}}\right) - 1\right]
W(p_1,p_2\rightarrow p_1',p_2')
\label{div-S}
\end{equation}
(note that $\epsilon = \epsilon^{-1}$ for $\epsilon = \pm1$).
Eqs.\ \eqref{entropy-density-current} and \eqref{div-S} reduce to
Eqs.\ \eqref{entropy-density-current_MB} and \eqref{div-S_MB} in the limit $\epsilon \to 0$,
and thus they actually cover all three cases $\epsilon = 0$, $\pm 1$.

Now the (real) function $f(z) = z - 1 - \ln z$ is positive for $z > 0$, $z \ne 1$,
taking its absolute minimum zero at $z = 1$.
Therefore, as long as all the phase space factors $\tilde{p}_{i0}$ and $\tilde{p}'_{i0}$
are positive, the right hand side of Eq.\ \eqref{div-S} is non-negative.
Integrating over three-space, we can write
\begin{equation}
\int d^3x \,\partial_\mu s^\mu =
\int d^3x\left(\partial_t s^0 + \vec{\nabla}\cdot\vec{s}\right) =
\frac{d}{dt}\left(\int d^3 x\,s^0\right) +
\int_{s_{r\to\infty}}d^2\sigma(\hat n\cdot \vec{s})\>.
\end{equation}
For a localized distribution the last term tends to zero,
and thus we conclude that the total entropy
\begin{equation}
S = \int d^3x\, s^0 
= \frac{kg}{h^3}\int d^6\xi\left[\bigl(\epsilon^{-1} + f(x,p)\bigr)\ln\bigl(1 + \epsilon f(x,p)\bigr)
- f(x,p)\ln f(x,p)\right]
\label{entropy}
\end{equation}
will never decrease.

Moreover, it follows that $S$ will be stationary iff
$z = f_{1'} f_{2'}\bar{f}_1\bar{f}_2/(f_1 f_2\bar{f}_{1'}\bar{f}_{2'}) = 1$, or,
defining $\phi(x,p) = -\ln \bigl(f(x,p)/\bar{f}(x,p)\bigr)$,
\begin{equation}
\phi_{1'} + \phi_{2'} - \phi_1 -\phi_2 = 0\>.
\label{equilibrium-condition}
\end{equation} 
Now note that the momenta $p_1^\mu$, $p_2^\mu$, $p'_1{}^{\mu}$ and $p'_2{}^{\mu}$
satisfy the constraint
\begin{equation}
p_1^\mu + p_2^\mu - p'_1{}^{\mu} - p'_2{}^{\mu} = 0\>.
\label{momentum-conservation}
\end{equation}
It can be shown \cite{CercignaniKremer} that the most general solution
of the equilibrium condition \eqref{equilibrium-condition} with momenta
constrained by condition \eqref{momentum-conservation} 
(amounting to a so-called \textit{summational invariant)} is given by
\begin{equation}
\phi(x,p) = -\alpha(x) + \beta_\mu(x)p^\mu 
\quad \Rightarrow \quad
f(x,p) = \frac{1}{e^{-\alpha(x) + \beta_\mu(x)p^\mu} - \epsilon} \>.
\label{Boltzmann-Juttner-form}
\end{equation}
for arbitrary scalar function $\alpha(x)$ and four-vector function $\beta_\mu(x)$. 

The equilibrium solution has to satisfy the Uehling-Uhlenbeck equation \eqref{UU-eq}.
As the collision term is guaranteed to vanish, the left-hand side
of the Uehling-Uhlenbeck equation has to vanish as well.
Let us consider the case $K^\mu = 0$ in which there is no external force.
It then follows that $\alpha(x)$ and $\beta_\mu(x)$ have to satisfy
\begin{equation}
\tilde{p}^\mu \partial_\mu\alpha - \tilde{p}^\mu p^\nu\partial_\mu\beta_\nu = 0\>,
\label{conditions-alpha-beta}
\end{equation}
where $\tilde{p}^\mu$ is given by Eq.\ \eqref{tilde-p}.
Eq.\ \eqref{conditions-alpha-beta} can be written as
\begin{equation}
\tilde{p}^\mu\left(\partial_\mu\alpha -
\partial_\mu\beta_\kappa\bigl((\delta+c)^{-1}\bigr)^\kappa{}_\lambda a^\lambda\right)
- \tilde{p}^\mu \tilde{p}^\nu\,\partial_\mu \beta_\kappa
\bigl((\delta+c)^{-2}\bigr)^\kappa{}_\nu = 0\>.
\label{conditions-alpha-beta2}
\end{equation}
As this identity has to be satisfied for arbitrary momentum,
it follows that both terms of Eq.\ \eqref{conditions-alpha-beta2} have to vanish
independently.
The second term yields the Killing equation
\begin{equation}
\partial_\mu\tilde{\beta}_\nu + \partial_\nu\tilde{\beta}_\mu = 0
\end{equation}
with $\tilde{\beta}_\mu = \beta_\kappa\bigl((\delta+c)^{-2}\bigr)^\kappa{}_\mu$.
It can be shown that the most general solution of this condition is
\cite{Weinberg}
\begin{equation}
\tilde{\beta}_\mu(x) = \omega_{\mu\nu} x^\nu + \tilde{\beta}_{\mu 0}
\label{tilde-beta-mu}
\end{equation}
where $\tilde{\beta}_{\mu0}$ is an arbitrary constant four-vector,
and the constant two-tensor $\omega_{\mu\nu}$ satisfies
\begin{equation}
\omega_{\mu\nu} = -\omega_{\nu\mu}\>.
\label{condition-omega}
\end{equation}
Next, by substituting the result \eqref{tilde-beta-mu} into the first term
of Eq.\ \eqref{conditions-alpha-beta2} and imposing it to vanish,
it follows that the most general solution for $\alpha(x)$ is
\begin{equation}
\alpha(x) = \alpha - x^\mu \omega_{\mu\lambda}(\delta+c)^\lambda{}_\nu a^\nu
\end{equation}
where $\alpha$ is a constant.
For a time-independent distribution we must have
$\omega_{0\mu} = 0$.
Then the $\omega$-dependent terms represent a fluid which is rigidly rotating.
In the following we will assume $\omega_{\mu\nu} = 0$, and we obtain for
the equilibrium distribution
\begin{equation}
f_{\mbox{\tiny eq}}(x,p) =\frac{1}{e^{-\alpha + \theta_\mu p^\mu} - \epsilon} 
\label{equilibrium-distribution}
\end{equation}
with $\theta_\mu = \tilde{\beta}_{\lambda0}\bigl((\delta+c)^{-2}\bigr)^\lambda{}_\mu$,
which is now independent of $x$.

\section{The particle number current and the energy-momentum tensor}
\label{sec:particle-current_EM-tensor}

We obtain the particle number density by integrating $g.f(x,p)/h^3$ over all three-momenta:
\begin{equation}
J^0(x) = \frac{g}{h^3}\int d^3p\,f(x,p)\>,
\label{J0}
\end{equation}
while the particle number current density is given by
\begin{equation}
J^i(x) = \frac{g}{h^3}\int d^3p\,v^i f(x,p)\>. 
\label{Ji}
\end{equation}
Combining expressions \eqref{J0} and \eqref{Ji} we obtain the observer Lorentz-covariant expression
\begin{equation}
J^\mu(x) = \frac{g}{h^3}\int \frac{d^3p}{\tilde{p}^0}\tilde{p}^\mu f(x,p)
\label{Jmu}
\end{equation}
where we used Eq.\ \eqref{vi} defining the particle number four-current density.
It can be checked that it is conserved, $\partial_\mu J^\mu(x) = 0$,
by taking $\psi(x,p) = 1$ in the transfer equation \eqref{transfer-equation}.
Note that its space-like part involves the Lorentz-violating coefficients $c^{\mu\nu}$ and $a^\mu$.

Similarly, we define the components of the canonical energy-momentum tensor as
\begin{align}
T^{00}(x) &= \frac{g}{h^3}\int d^3p\,p^0 f(x,p) \qquad \text{(energy density)}\\
T^{0i}(x) &= \frac{g}{h^3}\int d^3p\,p^i f(x,p) \qquad \text{(momentum density)}\\
T^{i0}(x) &= \frac{g}{h^3}\int d^3p\,v^i p^0 f(x,p) \qquad \text{(energy density flow)}\\
T^{ij}(x) &= \frac{g}{h^3}\int d^3p\,v^i p^j f(x,p) \qquad \text{(momentum density flow)}
\end{align}
which can be combined in the manifestly Lorentz-covariant expression
\begin{equation}
T^{\mu\nu}(x) = \frac{g}{h^3}\int \frac{d^3p}{\tilde{p}^0}\tilde{p}^\mu p^\nu f(x,p)
\label{Tmunu}
\end{equation}
upon using expression \eqref{vi}.
Note that it is nonsymmetric, $T^{\mu\nu} \ne T^{\nu\mu}$,
because the velocity four-vector $\tilde{p}^\mu/m$ is not proportional to the
four-momentum $p^\mu$,
due to the presence of the Lorentz-violating coefficients $c^{\mu\nu}$ and $a^\mu$.
By taking $\psi(x,p) = p^\nu$ in the transfer equation \eqref{transfer-equation}
one obtains $\partial_\mu T^{\mu\nu} = 0$.
That is, $T^{\mu\nu}$ is conserved (in the absence of external forces!).
However, note that, in general, $\partial_\nu T^{\mu\nu} \ne 0$!

The appearance of the nonsymmetric energy-momentum tensor \eqref{Tmunu}
is not a surprise.
It is well known that the presence of Lorentz-violating coefficients
in quantum field theory gives rise
to nonsymmetric canonical energy-momentum tensors \cite{ColladayKostelecky97,Kostelecky03}.
\\

Let us try to evaluate the current density and the energy momentum tensor for the
equilibrium distribution \eqref{equilibrium-distribution}.
To this effect, we define the generating function
\begin{equation}
I = \int \frac{d^3p}{\tilde{p}^0} \frac{1}{e^{-\alpha + \theta_\mu p^\mu} - \epsilon} \>.
\label{generating-function}
\end{equation}
It then follows that
\begin{align}
\label{J-I}
\frac{\partial J^\mu}{\partial\alpha}  &= -\frac{g}{h^3}(\eta + c)^{\mu\lambda}
\left((\delta + c)_\lambda{}^\nu\frac{\partial}{\partial\theta^\nu} + a_\lambda\frac{\partial}{\partial\alpha}\right)I \\
\frac{\partial T^{\mu\nu}}{\partial\alpha} &=  -\frac{\partial J^\mu}{\partial\theta_\nu}\>.
\label{T-I-J}
\end{align}
We can evaluate the integral in Eq.\ \eqref{generating-function} by introducing
\begin{equation}
\bar{p}\,^\mu = (\delta+c)^\mu{}_\lambda p^\lambda - a^\mu 
\quad\Leftrightarrow\quad
p^\mu = \bigl((\delta+c)^{-1}\bigl)^\mu{}_\alpha(\bar{p} + a)^\alpha\>.
\end{equation}
The dispersion relation \eqref{dispersion-relation} then becomes
\begin{equation}
\bar{p}\cdot\bar{p} - m^2 = 0
\label{dispersion-relation_bar-p}
\end{equation}
and we can write
\begin{align}
I &= 2\int d^4p\,\delta\left(\bigl((\delta+c)^\mu{}_\lambda p^\lambda - a^\mu\bigr)
\bigl((\delta+c)_\mu{}^\nu p_\nu - a_\mu\bigr) - m^2\bigr)\right)
\frac{\theta\bigl((\delta+c)^0{}_\lambda p^\lambda - a^0\bigr)}{e^{-\alpha + \theta_\mu p^\mu} - \epsilon} \nonumber\\
 &= \frac{2}{|\,\delta + c\,|} \int d^4\bar{p}\,\theta(\bar{p}^0)\delta(\bar{p}\cdot\bar{p} - m^2)
\frac{1}{e^{-\alpha + \theta_\mu p^\mu} - \epsilon} \nonumber\\
&= \frac{2}{|\,\delta + c\,|} \int d^4\bar{p}\,\theta(\bar{p}^0)\delta(\bar{p}\cdot\bar{p} - m^2)
\frac{1}{e^{-\alpha + \tilde{\theta}\cdot(\bar{p}+a)} - \epsilon}
\label{I-reps}
\end{align}
where
$\tilde{\theta}^\mu = \theta^\nu\bigl((\delta+c)^{-1}\bigr)_\nu{}^\mu$,
$|\delta + c|$ denotes the determinant of the matrix $(\delta + c)^\mu{}_\nu$
and $\theta(\bar{p}^0)$ the Heaviside function of $\bar{p}^0$.
Going to a frame in which $\tilde{\theta}^\mu$ is purely timelike, i.e.,
\begin{equation}
\tilde{\theta}^\mu = (\bar{\theta}; \vec{0})\>,\qquad 
\bar{\theta} = \sqrt{\tilde{\theta}\cdot\tilde{\theta}}
= \sqrt{\theta_\mu\bigl((\delta+c)^{-2}\bigr)^\mu{}_\nu \theta^\nu}\>,
\label{bar-theta}
\end{equation}
it follows, upon parametrizing $|\vec{\bar{p}}\,| = m\sinh x$, that
\begin{equation}
I = \frac{4\pi m^2}{|\,\delta + c\,|} \int_0^\infty \frac{\sinh^2 x\,dx}{e^{-\alpha + \tilde{\theta}\cdot a + \bar{\theta}\cosh x} - \epsilon}
= \frac{4\pi m^2}{|\,\delta + c\,|} J_{20}(m\bar{\theta},\alpha - \theta\cdot\tilde{a}) 
\label{I-rep-x}
\end{equation}
where
\begin{equation}
\tilde{a}^\mu = \bigl((\delta+c)^{-1}\bigr)^\mu{}_\lambda a^\lambda\>,
\label{tilde-a}
\end{equation}
and 
\begin{equation}
J_{nm}(\zeta,\alpha) = \int_0^\infty \frac{\sinh^n x \cosh^m x}{e^{\zeta\cosh x - \alpha} - \epsilon} dx \>.
\label{Jnm}
\end{equation}
Note that we have assumed here that the mass $m$ is nonzero.
The massless case will be treated separately in section \ref{sec:massless}.
The functions $J_{nm}(\zeta,\alpha)$ satisfy the identities
\begin{align}
\label{id-J-alpha}
\frac{\partial J_{nm}}{\partial\alpha} &= \frac{n-1}{\zeta} J_{n-2,m+1} + \frac{m}{\zeta} J_{n,m-1} \qquad (n\ge2)\\
\frac{\partial J_{nm}}{\partial\zeta} &= -\frac{n-1}{\zeta} J_{n-2,m+2} - \frac{m+1}{\zeta} J_{nm}
= - \frac{\partial J_{n,m+1}}{\partial\alpha}\>.
\label{id-J-zeta}
\end{align}
By using Eqs.\ \eqref{id-J-alpha} and \eqref{id-J-zeta} one readily finds
\begin{equation}
\frac{\partial I}{\partial\theta^\mu} = -\frac{4\pi m^2}{|\,\delta + c\,|}\frac{\partial}{\partial\alpha}
\left[\frac{m}{\bar{\theta}} J_{21}(m\bar{\theta},\alpha-\theta\cdot\tilde{a})\,\theta_\nu\bigl((\delta+c)^{-2}\bigr)^\nu{}_\mu
- J_{20}(m\bar{\theta},\alpha-\theta\cdot\tilde{a}) \tilde{a}_\mu\right]\>.
\end{equation}
From identity \eqref{J-I} it then follows after integrating over the parameter $\alpha$ that
\begin{equation}
J^\mu = \frac{4\pi g m^3}{h^3\bar{\theta}|\,\delta + c\,|} J_{21}(m\bar{\theta},\alpha-\theta\cdot\tilde{a}) \theta^\mu\>.
\label{J_formula}
\end{equation}
Note that there is no additional term independent of $\alpha$
(i.e., integration constant) as both of the expressions
\eqref{generating-function} and \eqref{Jmu} vanish in the $\alpha \to -\infty$ limit.

Similarly, it follows from Eqs.\ \eqref{T-I-J}, \eqref{id-J-alpha} and \eqref{id-J-zeta} that
\begin{equation}
T^{\mu\nu} = \frac{4\pi g m^4}{h^3 |\,\delta + c\,|}
\left[-\frac{1}{3}J_{40}\eta^{\mu\nu} + \left(J_{22} + \frac{1}{3}J_{40}\right)
\frac{\theta^\mu\theta^\lambda\bigl((\delta+c)^{-2}\bigr)_\lambda{}^\nu}{\bar{\theta}^2}
+ J_{21}\frac{\theta^\mu\tilde{a}^\nu}{m\bar{\theta}}\right]\>,
\label{T_formula}
\end{equation}
where we abbreviated $J_{nm}(m\bar{\theta},\alpha-\theta\cdot\tilde{a}) = J_{nm}$.

Defining the observer scalar quantity
\begin{equation}
n = \sqrt{J^\mu\bigl((\delta+c)^{-2}\bigr)_\mu{}^\nu J_\nu} = 
\frac{4\pi g m^3}{h^3 |\,\delta + c\,|} J_{21}(m\bar{\theta},\alpha - \theta\cdot\tilde{a})
\label{n}
\end{equation}
it follows from Eq.\ \eqref{J_formula} that
\begin{equation}
J^\mu = \frac{n}{\bar{\theta}}\,\theta^\mu
\label{Jmu_eq}
\end{equation}
and from Eq.\ \eqref{T_formula} that
\begin{equation}
T^{\mu\nu} = -\frac{n}{\bar{\theta}}
\left[\frac{m\bar{\theta}J_{40}}{3J_{21}}\eta^{\mu\nu}
- \frac{J_{22}+\tfrac13 J_{40}}{J_{21}} \frac{m}{\bar{\theta}}
\theta^\mu \theta^\lambda\bigl((\delta+c)^{-2}\bigr)_\lambda{}^\nu
- \theta^\mu\tilde{a}^\nu \right]\>.
\label{Tmunu_eq}
\end{equation}
It is useful to define the four-vector
\begin{equation}
u^\mu = \frac{J^\mu}{n} = \frac{\theta^\mu}{\bar{\theta}}\>.
\label{u-mu}
\end{equation}
It satisfies the normalization condition
\begin{equation}
u^\mu\bigl((\delta+c)^{-2}\bigr)_\mu{}^\nu u_\nu = 1\>.
\label{u-normalization}
\end{equation}
Comparison with Eq.\ \eqref{constraint} (in the gauge $e = 1/m$)
shows that $u^\mu$ can be interpreted as the four-velocity of the fluid.
Observer frames in which the spacelike components of $u^\mu$ are identically zero
define the rest frame of the fluid.
For such a frame, the time component of $u^\mu$ is equal to
\begin{equation}
u^0 = \left(\bigl((\delta+c)^{-2}\bigr)_0{}^0\right)^{-1/2}
\label{u0}
\end{equation}
(rather than one, in the Lorentz-symmetric case).
In the rest frame of the fluid,
we can identify $\theta^0$, the time component of $\theta^\mu$,
with $(kT)^{-1}$. 
It then follows from Eqs.\ \eqref{u-mu} and \eqref{u0} that
\begin{equation}
\bar{\theta} = \frac{1}{u^0 kT} = \frac{\sqrt{\bigl((\delta+c)^{-2}\bigr)_0{}^0}}{kT}\>.
\label{bar-theta_T}
\end{equation}
The latter expression is valid in any frame.

In order to extract the physically relevant quantities from the energy-momentum tensor,
such as the pressure and the energy density,
one defines projective tensors $\Delta_\parallel^{\mu\nu}$ and
$\Delta_\perp^{\mu\nu} = \eta^{\mu\nu} - \Delta_\parallel^{\mu\nu}$ such that
\begin{align}
\label{projector-id-1}
\Delta_\parallel^{\mu\beta}\Delta_{\parallel,\beta}{}^\nu & = \Delta_\parallel^{\mu\nu} \\
\Delta_\perp^{\mu\beta}\Delta_{\perp,\beta}{}^\nu & = \Delta_\perp^{\mu\nu} \\
\Delta_\parallel^{\mu\beta}\Delta_{\perp,\beta}{}^\nu & = \Delta_\perp^{\mu\beta}\Delta_{\parallel,\beta}{}^\nu = 0 \\
\Delta_\parallel^{\mu\nu} u_\nu &= u^\mu  \\
\Delta_\perp^{\mu\nu} u_\nu &= 0\>.
\label{projector-id-5}
\end{align}
One then defines the energy density in the 
rest frame of the fluid as
\begin{equation}
\mathcal{E} = \Delta_\parallel^{\nu\mu} T_ {\mu\nu}\>,
\label{energy-T}
\end{equation}
the pressure as
\begin{equation}
P = -\frac{1}{3}\Delta_\perp^{\nu\mu} T_ {\mu\nu}\>,
\label{P-T}
\end{equation}
and the energy-momentum tensor can be written, in any frame, as
\begin{equation}
T^{\mu\nu} = \mathcal{E} \Delta_\parallel^{\mu\nu} - P \Delta_\perp^{\mu\nu}\>.
\label{T-decomposition}
\end{equation}

In the Lorentz-symmetric case one simply has $\Delta_\parallel^{\mu\nu} = u^\mu u^\nu$.
For the Lorentz-violating case the bi-vector $u^\mu u^\nu$ no longer satisfies the identities
\eqref{projector-id-1}--\eqref{projector-id-5}.
It turns out that the appropriate generalization satisfying identities
\eqref{projector-id-1}--\eqref{projector-id-5}
and also yielding the correct decomposition \eqref{T-decomposition} is given by
\begin{equation}
\Delta_\parallel^{\mu\nu} = \frac{u^\mu\left[m (J_{22} + \tfrac13 J_{40})\bigl((\delta+c)^{-2}\bigr)^\nu{}_\lambda u^\lambda + J_{21}\tilde{a}^\nu\right]}{m(J_{22}+\tfrac13 J_{40}) + (\tilde{a}.u)J_{21}}
\end{equation}
It follows from Eqs.\ \eqref{energy-T} and \eqref{P-T} that
\begin{align}
\label{E}
\mathcal{E} &= nm\left(\frac{J_{22}}{J_{21}} + \frac{\tilde{a}\cdot u}{m}\right)\\
P &= \frac{nm}{3} \frac{J_{40}}{J_{21}}\>.
\label{P-n-theta}
\end{align}

Next consider the entropy density current \eqref{entropy-density-current}
for the equilibrium distribution \eqref{equilibrium-distribution}.
A little algebra shows that
\begin{align}
s^\mu &= \frac{kg}{h^3}\int \frac{d^3p}{\tilde{p}_0} \tilde{p}^\mu 
\left[-\frac{1}{\epsilon}\ln\bigl(1-\epsilon\,e^{\alpha - \theta\cdot p}\bigr)
+ \frac{\theta\cdot p - \alpha}{e^{\theta\cdot p - \alpha} - \epsilon}\right]\nonumber\\
&= s_1^\mu + s_2^\mu + s_3^\mu\>.
\label{Smu_eq}
\end{align}
where
\begin{align}
s_1^\mu &= -\frac{kg}{\epsilon h^3}\int \frac{d^3p}{\tilde{p}_0} \tilde{p}^\mu 
\ln\bigl(1 - \epsilon e^{\alpha-\theta\cdot p}\bigr) \\
\label{S2mu}
s_2^\mu &= \frac{kg}{h^3}\int \frac{d^3p}{\tilde{p}_0} \tilde{p}^\mu 
\frac{\theta\cdot p}{e^{\theta\cdot p-\alpha} - \epsilon} = kT^{\mu\nu}\theta_\nu \\
s_3^\mu &= -\frac{k\alpha g}{\epsilon h^3}\int \frac{d^3p}{\tilde{p}_0} \tilde{p}^\mu 
\frac{1}{e^{\theta\cdot p-\alpha} - \epsilon} = -k\alpha J^\mu
\label{S3mu}
\end{align}
In order to evaluate $s_1^\mu$ we note that
\begin{equation}
\frac{\partial s_1^\mu }{\partial\alpha} = -\frac{1}{\alpha}s_3^\mu = kJ^\mu \>.
\label{dS1-dalpha_Jmu}
\end{equation}
From Eq.\ \eqref{id-J-alpha} we have
\begin{equation}
\frac{\partial J_{40}(\zeta,\alpha)}{\partial\alpha} = \frac{3}{\zeta}J_{21}(\zeta,\alpha)
\end{equation}
which allows to express Eq.\ \eqref{J_formula} as
\begin{equation}
J^\mu = \frac{4\pi g m^4}{3h^3 |\,\delta + c\,|}
\frac{\partial J_{40}(m\bar{\theta},\alpha-\theta\cdot\tilde{a})}{\partial\alpha} \theta^\mu\>.
\end{equation}
It then follows from Eq.\ \eqref{dS1-dalpha_Jmu} after integrating over $\alpha$ that
\begin{equation}
s_1^\mu = \frac{4\pi kg m^4}{3h^3|\,\delta + c\,|}
J_{40}(m\bar{\theta},\alpha-\theta\cdot \tilde{a}) \theta^\mu\>.
\label{S1mu}
\end{equation}
Note that there is no term independent of $\alpha$ (integration constant) as both $s_1^\mu$ and $J_{40}$ vanish
in the $\alpha\to\infty $ limit.

We can compute $s_2^\mu$ by evaluating $T^{\mu\nu}\theta_\mu$ from Eqs.\ \eqref{Tmunu_eq}
and \eqref{u-mu}. It follows that
\begin{equation}
T^{\mu\nu}\theta_\mu = n\left(m\frac{J_{22}}{J_{21}} + \tilde{a}\cdot u\right)\theta^\mu
\label{T-theta}
\end{equation}
From Eqs.\ \eqref{Smu_eq}--\eqref{S3mu}, \eqref{Jmu_eq} one finds
\begin{align}
s^\mu &= kn\left(m\frac{J_{22} + \tfrac13 J_{40}}{J_{21}} + \tilde{a}\cdot u\right)\theta^\mu - k\alpha nu^\mu\nonumber\\
&= k\bar{\theta}(\mathcal{E} + P)u^\mu - k\alpha n u^\mu\>,
\label{entropy-density-current-eq}
\end{align}
where we used Eqs.\ \eqref{energy-T} and \eqref{P-T}. 
Considering Eq.\ \eqref{entropy-density-current-eq} in the rest frame,
we can write the time component as
\begin{equation}
\alpha kT = \frac{\mathcal{E} + P - T s^0}{nu^0}\>.
\label{Gibbs-Duhem}
\end{equation}
On the right-hand side of Eq.\ \eqref{Gibbs-Duhem} we recognize the Gibbs function
(free energy) per particle
(note that $nu^0 = J^0$ is equal to the particle density in the rest frame).
Relation \eqref{Gibbs-Duhem} corresponds to the Gibbs-Duhem equation for one type
of particle, and we can identify
\begin{equation}
\alpha = \frac{\mu_E}{kT}\>.
\end{equation}
where $\mu_E$ is the chemical potential.

\section{The Maxwell-Boltzmann case}
\label{sec:Maxwell-Boltzmann}

For either of the cases $\epsilon = \pm1$ the above results simplify,
in the limit $e^{-\alpha} = e^{-\mu_E/(kT)} \gg 1$,
to the Maxwell-Boltzmann case $\epsilon = 0$.
The equilibrium distribution function \eqref{equilibrium-distribution} then
simplifies to the Maxwell-J\"uttner distribution
\begin{equation}
f_{\mbox{\tiny eq}}(x,p) = e^\alpha.e^{-\theta_\mu p^\mu}
\label{equilibrium-distribution_MB}
\end{equation}
while the functions $J_{nm}(\zeta,\alpha)$ defined in Eq.\ \eqref{Jnm} reduce to
\begin{equation}
J_{nm}(\zeta,\alpha) \to e^\alpha \int_0^\infty \sinh^n x \cosh^m x e^{-\zeta\cosh x} dx\>.
\label{Jnm_MB}
\end{equation}
These functions can be related the modified Bessel functions of the second kind
which have the representation
\begin{equation}
K_n(\zeta) = \int_0^\infty \cosh(nx) e^{-\zeta \cosh x} dx\>.
\end{equation}
They satisfy the recurrence relation
\begin{equation}
K_{n+1}(\zeta) - K_{n-1}(\zeta) = \frac{2n}{\zeta} K_{n}(\zeta)\>.
\label{K-recurrence}
\end{equation}
By using relation \eqref{K-recurrence}, together with the identities
\begin{align}
\sinh^2 x \cosh x &= \tfrac14 \bigl(\cosh(3x) - \cosh x\bigr) \\
\sinh^2 x \cosh^2 x &= \tfrac18 \bigl(\cosh(4x) - 1\bigr) \\
\sinh^4 x &= \tfrac18\bigl(\cosh(4x) - 4\cosh(2x) + 3\bigr)
\end{align}
it is then straightforward to show that, in the Maxwell-Boltzmann limit,
\begin{align}
J_{21}(\zeta,\alpha) &\to \frac{e^\alpha}{\zeta} K_2(\zeta)\\
J_{22}(\zeta,\alpha) &\to \frac{e^\alpha}{4\zeta} \bigl(3 K_3(\zeta) + K_1(\zeta)\bigr)\\
J_{40}(\zeta,\alpha) &\to \frac{3e^\alpha}{\zeta^2} K_2(\zeta)\>.
\end{align}
The expressions \eqref{n} for $n$, \eqref{E} for $\mathcal{E}$ and \eqref{P-n-theta} for $P$ reduce to
\begin{align}
\label{n_MB}
n &= \frac{4\pi g m^2}{h^3 |\,\delta + c\,|}\,e^{\alpha - \theta\cdot\tilde{a}}
\frac{K_2(m\bar{\theta})}{\bar{\theta}} \\
\label{E_MB}
\mathcal{E} &= n\left(m G(m\bar{\theta}) - \frac{1}{\bar{\theta}} + \tilde{a}\cdot u\right) \\
\label{P-n-theta_MB}
P &= \frac{n}{\bar{\theta}}\>.
\end{align}
Here we followed \cite{CercignaniKremer} in defining $G(u) = K_3(u)/K_2(u)$.
It can be checked that relation \eqref{n_MB} is consistent
with Eqs.\ (31) and (27) of Ref.\ \cite{ColladayMcDonald},
if we identify $\langle N^{(C)}\rangle/V$ with our $J^0$ and take $g = 2$
(as Ref.\ \cite{ColladayMcDonald} considers a spin 1/2 system)
and take into account that the chemical potential considered in Ref.\ \cite{ColladayMcDonald}
does not include the rest energy of the particle.

In the rest frame we can write Eq.\ \eqref{P-n-theta_MB} as $P = J^0/\theta^0$.
As in that frame $J^0$ can be identified with the particle density
and $\theta^0$ with the inverse temperature,
Eq.\ \eqref{P-n-theta_MB} thus amounts to the ideal gas law.

Eq.\ \eqref{E_MB} can be used to determine the specific heat per particle in the 
rest frame.
As the energy per particle equals $e = \mathcal{E}/J^0 = \mathcal{E}/(n u^0)$,
it is straightforward to compute the specific heat per particle at constant volume (i.e., at constant $n$):
\begin{align}
c_V &= \left(\frac{\partial e}{\partial T}\right)_n 
= \frac{1}{u^0n}\frac{\partial\bar{\theta}}{\partial T}\left(\frac{\partial\mathcal{E}}{\partial \bar{\theta}}\right)_n \nonumber\\
&= -k\bar{\theta}^2\left(m^2 G'(m\bar{\theta}) + \frac{1}{\bar{\theta}^2}\right) \nonumber\\
&= k\left[(m\bar{\theta})^2\bigl(1 - G(m\bar{\theta})^2\bigr) + 5m\bar{\theta}\, G(m\bar{\theta}) - 1\right]
\label{cV}
\end{align}
where we used the identity $G'(u) = G(u)^2 - 5G(u)/u - 1$.
Note that $c_V$ does not depend on the LV parameter $a^\mu$.
Its only difference from the Lorentz-symmetric result (see \cite{CercignaniKremer}) is through
the replacement $T \to (k\bar{\theta})^{-1} = u^0 T$.
One can also compute the specific heat per particle at constant pressure by
introducing the enthalpy per particle $h = e + P/J^0$.
One obtains $c_P = c_V + k$ as usual,
essentially as a consequence of the ideal gas law \eqref{P-n-theta_MB}.
The low-temperature limit $\bar{\theta} \to \infty$ yields
the familiar results $c_V \to 3k/2$ and $c_P \to 5k/2$,
while in the ultrarelativistic limit $\bar{\theta} \to 0$,  $c_V \to 3k$ and $c_P \to 4k$.

Expression \eqref{Tmunu_eq} for the energy-momentum tensor becomes
\begin{equation}
T^{\mu\nu} = \frac{n}{\bar{\theta}}\left(\frac{m}{\bar\theta}G(m\bar{\theta})
\theta^\mu\bigl((\delta+c)^{-2}\bigr)^\nu{}_\lambda \theta^\lambda
-\eta^{\mu\nu} + \theta^\mu \tilde{a}^\nu\right)\>,
\end{equation}
while with the use of Eq.\ \eqref{P-n-theta_MB} the entropy current
density \eqref{entropy-density-current-eq} simplifies to
\begin{equation}
s^\mu = k\bigl[\bar{\theta}\mathcal{E} + (1-\alpha)n\bigr] u^\mu \>.
\label{entropy-density-current-eq_MB}
\end{equation}

We can use expression \eqref{n_MB} to express the equilibrium value
of the parameter $\alpha$ in terms of the particle density.
It follows that
\begin{equation}
\alpha = \ln\left(\frac{|\,\delta + c\,|\, n \bar{\theta}h^3}{4\pi g m^2 K_2(m\bar{\theta})}\right) + \theta\cdot \tilde{a}
\label{alpha_MB}
\end{equation}
which means that the Maxwell-Boltzmann equilibrium distribution \eqref{equilibrium-distribution_MB}
can be written as
\begin{equation}
f_{\mbox{\tiny eq}}(x,p) = \frac{|\,\delta + c\,|\, n \bar{\theta}h^3}{4\pi g m^2 K_2(m\bar{\theta})}
e^{-\theta\cdot(p - \tilde{a})}
\label{equilibrium-distribution_MB_2}
\end{equation}
Using Eqs.\ \eqref{alpha_MB}, \eqref{E_MB} and \eqref{entropy-density-current-eq_MB}
we then find for the entropy current density
\begin{equation}
s^\mu = kn \left[m\bar{\theta}G(m\bar{\theta})
- \ln\left(\frac{|\,\delta + c\,|\, n \bar{\theta}h^3}{4\pi g m^2 K_2(m\bar{\theta})}\right)\right]u^\mu\>.
\label{entropy-density-current-eq_MB}
\end{equation}
Note that the right-hand side of Eq.\ \eqref{entropy-density-current-eq_MB} is independent of $a^\mu$;
this is not the case for the corresponding expression in Eq.\ \eqref{entropy-density-current-eq}
which still depends on $a^\mu$ through the arguments of the functions $J_{nm}$ in the definitions
\eqref{E} and \eqref{P-n-theta} of $\mathcal{E}$ and $P$.

\subsection{Nonrelativistic limit}
Let us now consider the nonrelativistic limit, in which the thermal kinetic energy of the particles
is much less than their rest mass.
In this case, we have that $\zeta \equiv m\bar{\theta} \gg 1$, and we can use the large-$\zeta$ expansion
\cite{Gradsteyn}:
\begin{equation}
K_n(\zeta) = \sqrt{\frac{\pi}{2\zeta}}e^{-\zeta}\left[1 + \frac{4n^2-1}{8\zeta} + \frac{(4n^2-1)(4n^2-9)}{2!(8\zeta)^2} + \ldots\right]\>.
\end{equation}
It follows that
\begin{equation}
G(\zeta) = \frac{K_3(\zeta)}{K_2(\zeta)} = 1 + \frac{5}{2\zeta} + \frac{15}{8\zeta^2} + \ldots
\end{equation}
and we get for the energy per particle in the rest frame
\begin{align}
e = \frac{\mathcal{E}}{u^0 n} &= \frac{1}{u^0}\left(m G(m\bar{\theta}) - \frac{1}{\bar{\theta}} + \tilde{a}\cdot u \right)\nonumber\\
&= \frac{m}{u^0} + \frac{3}{2}kT + \mathcal{O}\left(\frac{k^2T^2}{m}\right) + \tilde{a}^0\>.
\end{align}
The first term corresponds to the rest energy of the particle $m^*$,
which should be identified with its mass (note that we are taking natural units with $c = 1$).
Thus we have
\begin{equation}
m^* = \frac{m}{u^0} = m \sqrt{\bigl((\delta+c)^{-2}\bigr)^0{}_0}\>.
\label{effective-mass}
\end{equation}
From Eq.\ \eqref{entropy-density-current-eq_MB} we find
\begin{align}
\frac{s^\mu}{n} &= k\left[\zeta G(\zeta) - \ln\left(\frac{|\,\delta + c\,|\, n \bar{\theta}h^3}{4\pi g m^2 K_2(m\bar{\theta})}\right)\right]u^\mu\nonumber\\
&= k\left[\frac{5}{2} + \ln\left(\frac{g(2\pi m kTu^0)^{3/2}}{h^3|\,\delta + c\,|\,n}\right) + \frac{15}{4\zeta} + \ldots\right] u^\mu\nonumber\\
&= k\left[\frac{5}{2} + \ln\left(\frac{g(2\pi m^* kT)^{3/2}(u^0)^3}{h^3|\,\delta + c\,|\,n}\right) + \frac{15kT}{4m^*} + \ldots\right] u^\mu
\end{align}
The first two terms inside the square brackets correspond to the Sackur-Tetrode formula with Lorentz-violating correction;
the third term is the lowest-order relativistic correction.

\section{Massless particles}
\label{sec:massless}

Let us now consider the case $m = 0$ in which the particles are massless.
This will affect, first of all, the construction of the covariant Boltzmann equation.
In this case it is no longer possible to fix the reparametrization gauge by setting $e = 1/m$
as in Eq.\ \eqref{reparametrization-gauge}.
Instead, one can simply fix $e = 1$,
turning $\tau$ to be equal to an affine parameter of mass dimension $-2$ instead of $-1$.
Similarly, the four-force $K^\mu = dp^\mu/d\tau$ now becomes of mass dimension 3,
while the only change in the Boltzmann equation \eqref{Boltzmann-eq}
(as well as the Uehling-Uhlenbeck equation \eqref{UU-eq})
is the elimination of the factor $m$ in front of the force term.
The demonstration of the H-theorem as well as the derivation of the equilibrium distributions
in section \ref{sec:H-theorem} then go then go through practically unchanged.

Let us now analyze the evaluation of the particle number current and the 
energy-momentum tensor for the equilibrium distribution \eqref{equilibrium-distribution}.
We can no longer perform the parametrization $|\vec{\bar{p}}| = m \sinh x$
leading to Eq.\ \eqref{I-rep-x}.
Instead, we write, analogously to Eq.\ \eqref{I-reps},
\begin{align}
I &= \int \frac{d^3p}{\tilde{p}^0}\frac{1}{e^{-\alpha + \theta\cdot p} - \epsilon}\nonumber\\
&= 	\frac{2}{|\,\delta + c\,|} \int d^4p\,\theta(\bar{p}^0)\delta(\bar{p}\cdot\bar{p})
\frac{1}{e^{-\alpha + \tilde{\theta}\cdot(\bar{p} + a)} - \epsilon}\nonumber\\
&= \frac{1}{|\,\delta + c\,|}\int \frac{d^3\bar{p}}{\bar{p}^0}\frac{1}{e^{\alpha - \tilde{\theta}\cdot a + \tilde{\theta}\cdot\bar{p}} - \epsilon}\>.
\label{I-reps-massless}
\end{align}
The final integral can be carried out in the rest frame of $\tilde{\theta}^\mu$, yielding the result
\begin{equation}
I = \frac{4\pi\epsilon^{-1}}{|\,\delta + c\,|\,\bar{\theta}^2}\text{Li}_2\bigl(\epsilon\,e^{\alpha - \theta\cdot\tilde{\alpha}}\bigr)\>,
\end{equation}
where Li$_s(z)$ represents the polylogarithm function of order $s$, defined, for $|z| < 1$, by the power series
\begin{equation}
\text{Li}_s(z) = \sum_{k=1}^\infty \frac{z^k}{k^s}\>.
\end{equation}
With the help of relation \eqref{J-I} it follows, after integrating over $\alpha$, that
\begin{equation}
J^\mu = n\frac{\theta^\mu }{\bar{\theta}} = n u^\mu
\label{Jmu-massless}
\end{equation}
with 
\begin{equation}
n = \frac{8\pi g\epsilon^{-1}}{h^3|\,\delta + c\,|\,\bar{\theta}^3}\,
\text{Li}_3\bigl(\epsilon\,e^{\alpha - \theta\cdot\tilde{\alpha}}\bigr)
\label{n-massless}
\end{equation}
where we used the identity
\begin{equation}
\text{Li}_{s+1}(z) = \int_0^z\frac{\text{Li}_s(t)}{t}dt\>.
\label{Li-identity}
\end{equation}
Using relation \eqref{T-I-J} one then finds,
after once more integrating over $\alpha$ and using identity \eqref{Li-identity}, that
\begin{equation}
T^{\mu\nu} = -\frac{8\pi g\epsilon^{-1}}{h^3|\,\delta + c\,|\,\bar{\theta}^4}\left[
\text{Li}_4\bigl(\epsilon\,e^{\alpha - \theta\cdot\tilde{a}}\bigr)
\left(\eta^{\mu\nu} - 4\frac{\theta^\mu\bigl((\delta + c)^{-2}\bigr)^\nu{}_\lambda\theta^\lambda\bigl)}{\bar{\theta}^2}\right)
- \text{Li}_3\bigl(\epsilon\,e^{\alpha - \theta\cdot\tilde{a}}\bigr)\theta^\mu\tilde{a}^\nu\right]\>.
\label{Tmunu-masless}
\end{equation}
Next we construct the projectors $\Delta_\parallel^{\mu\nu}$ and $\Delta_\perp^{\mu\nu}$
satisfying the identities \eqref{projector-id-1}--\eqref{T-decomposition}.
We find
\begin{equation}
\Delta_\parallel^{\mu\nu} = \frac{u^\mu\left[4 K\bigl(\epsilon\,e^{\alpha - \theta\cdot\tilde{a}}\bigr)
\bigl((\delta + c)^{-2}\bigr)^\nu{}_\lambda u^\lambda + \bar{\theta}\,\tilde{a}^\nu\right]}
{4K\bigl(\epsilon\,e^{\alpha - \theta\cdot\tilde{a}}\bigr) + (\tilde{a}\cdot u)\bar{\theta}}\>,
\end{equation}
where we defined
\begin{equation}
 K(z) = \frac{\text{Li}_4(z)}{\text{Li}_3(z)}\>.
\end{equation}
We then find for the energy density
\begin{equation}
\mathcal{E} = \Delta_\parallel^{\nu\mu}T_{\mu\nu} =
n\left[\frac{3}{\bar{\theta}}\,K\bigl(\epsilon\,e^{\alpha - \theta\cdot\tilde{a}}\bigr) + \tilde{a}\cdot u\right]
\end{equation}
and the pressure
\begin{equation}
P = \frac{n}{\bar{\theta}}\,K\bigl(\epsilon\,e^{\alpha - \theta\cdot\tilde{a}}\bigr)\>.
\end{equation}
Using the decomposition \eqref{Smu_eq}--\ref{S3mu} we can compute the entropy density for the
massless case, with the use of expressions \eqref{Jmu-massless}, \eqref{n-massless} and \eqref{Tmunu-masless}.
It follows that
\begin{equation}
s^\mu = kn\left[4K\bigl(\epsilon\,e^{\alpha - \theta\cdot\tilde{a}}\bigr) + \tilde{a}\cdot\theta - \alpha\right]u^\mu\>.
\end{equation}
Note that the Gibbs-Duhem relation \eqref{Gibbs-Duhem} continues to hold.

Let us apply these results to a photon gas.
While the Boltzmann equation would seem to be inapplicable to this case,
as photons are essentially noninteracting,
they are nevertheless usually in thermal equilibrium,
due to the fact that they are in thermal contact with matter.
For this case we take $\alpha = 0$, $\epsilon = 1$ and $g = 2$ (corresponding to the two photon helicities).
Moreover, we will only assume a nonzero value for the Lorentz-violating coefficient $c^{\mu\nu}$.
To leading order, this coefficient equals $c^{\mu\nu} = -\frac12 (k_F)^{\mu\alpha\nu}{}_\alpha$,
where $k_F^{\mu\nu\rho\lambda}$ is the four-index Lorentz-violating coefficient in the 
photon sector of the minimal SME \cite{ColladayKostelecky98}.

It then follows from the identity Li$_4(1) = \pi^4/90$ that
\begin{equation}
\mathcal{E} =  aT^4
\end{equation}
with 
\begin{equation}
a = \frac{8\pi^5k^4(u^0)^4}{15h^3|\,\delta + c\,|}\>.
\end{equation}
This corresponds to the usual result accompanied by the Lorentz-violation correction
$(u^0)^4/|\,\delta + c\,|$.
The photon gas pressure, the entropy density and the energy density,
are related by $P = \mathcal{E}/3$ and $s^0 = 4\mathcal{E}/(3T)$,
just as in the Lorentz-symmetric case.

\section{Multiple species of particles}
\label{sec:multiple}

In the previous sections we considered the case in which there is only one
type of particle subject to the dispersion relation \eqref{dispersion-relation}.
In this case one may wonder whether the Lorentz-violating effects are really
physically measurable.
It is well known that in the SME the effects associated
with the $a^\mu$ coefficient on a single fermion field can be absorbed by a field redefinition
$\psi \to \exp\bigl(-ia_\mu x^\mu\bigr)\psi$ \cite{ColladayKostelecky98}.
Moreover, the symmetric $c^{\mu\nu}$ coefficient of the SME can be eliminated
by applying the linear transformation \cite{KosteleckyTasson09}
\begin{equation}
x^\mu \to (x')^\mu = \bigl(\delta^\mu_\nu + c^\mu{}_\nu\bigr)x^\nu
\label{linear-transformation}
\end{equation}
on the spacetime coordinates.%
\footnote{The transformation \eqref{linear-transformation} also affects the photon sector,
if present. For details, see Refs.\ \cite{KosteleckyBailey04,KosteleckyTasson09}.}

In this work we don't have a quantum field theory, but nevertheless,
the Lorentz-violating effects associated with the $c^{\mu\nu}$ coefficient
can be eliminated by applying the same spacetime transformation \eqref{linear-transformation}
in the defining particle Lagrangian \eqref{particle-action}.
Similarly, we can absorb the effect of the $a^\mu$ coefficient
by performing an appropriate momentum shift
redefining the zero point of energy momentum.
Clearly such an operation is compatible with momentum conservation
in the scattering process $(p_1, p_2 \to p_1', p_2')$.

In order to assure that the Lorentz-violating effects are physically observable
one needs to introduce other species of particles subject to different
Lorentz-violating sets of coefficients.
The spacetime transformation \eqref{linear-transformation}
can at most absorb the Lorentz-violating effects of the $c^{\mu\nu}$ coefficients
for one of the species of particles.
Similarly, we can only consistently apply a momentum shift to eliminate
the effects of the $a^\mu$ coefficients for one particle species.
Any differences between the coefficients of different species are unaffected
by either of the transformations.

Therefore, let us assume that there are $n > 1$ species of particles present,
labeled $a = 1,\ldots,n$.
Each species $a$ is defined by its mass $m_a$,
and by its own set of Lorentz-violating coefficients $c_a^{\mu\nu}$ and $a_a^\mu$.

First we will assume that only elastic collisions take place,
in which each particle maintains its identity.
Considering the Maxwell-Boltzmann case $\epsilon = 0$,
the Boltzmann equation \eqref{Boltzmann-eq} for species $a$ becomes
\begin{equation}
\tilde{p}_a^\mu\partial_\mu f_a + m\frac{\partial(K_a^\mu f_a)}{\partial p_a^\mu}
= \sum_{b=1}^n \frac{g}{h^3} \int\frac{d^3p_b}{\tilde{p}_{0b}}\frac{d^3p'_a}{\tilde{p}'_{0a}}\frac{d^3p'_b}{\tilde{p}'_{0b}}
(f_{a'} f_{b'} - f_{a} f_{b})W(p_a,p_b\rightarrow p_a',p_b')
\label{Boltzmann-eq-mult}
\end{equation}
and the corresponding transfer equation
\begin{equation}
\frac{\partial}{\partial x^\mu}\left( \int \frac{d^3p_a}{\tilde{p}_{0a}} f_a\psi_a\tilde{p}_a^\mu\right)
- \int \frac{d^3p_a}{\tilde{p}_{0a}}\left(\tilde{p}_a^\mu\frac{\partial\psi_a}{\partial x^\mu} + m K_a^\mu\frac{\partial\psi_a}{\partial p_a^\mu}\right) = \text{Coll}[\psi]\>,
\label{transfer-equation-mult}
\end{equation}
where now, applying detailed balance,
\begin{equation}
\text{Coll}[\psi]
=\frac{g}{2h^3} \sum_{b=1}^n \int \frac{d^3p_a}{\tilde{p}_{0a}}\frac{d^3p_b}{\tilde{p}_{0b}}
\frac{d^3p'_a}{\tilde{p}'_{0a}}\frac{d^3p'_b}{\tilde{p}'_{0b}} (\psi_{a'} + \psi_{b'} - \psi_a - \psi_b)
W(p_a,p_b\rightarrow p_a',p_b')f_a f_b\>.
\label{Coll-psi-mult}
\end{equation}
Now define the species-dependent currents and energy-momentum tensors
\begin{equation}
J_a^\mu(x) = \frac{g}{h^3}\int \frac{d^3p_a}{\tilde{p}_{0a}} \tilde{p}_a^\mu f_a(x,p_a)\>,\qquad
T_a^{\mu\nu}(x) = \frac{g}{h^3}\int \frac{d^3p_a}{\tilde{p}_{0a}} \tilde{p}_a^\mu p_a^\nu f_a(x,p_a)\>.
\end{equation}
Let us assume, as before, that the external force vanishes.
It then follows,
by taking $\psi = 1$ in Eq.\ \eqref{transfer-equation-mult}, that
\begin{equation}
\partial_\mu J_a^\mu = 0
\end{equation}
and by taking $\psi = p_a^\mu$ that
\begin{equation}
\partial_\mu T_a^{\mu\nu} = 0\>.
\end{equation}
In other words, for any species the particle number current
and the energy-momentum tensor are separately conserved.
It is easy to verify that the ansatz \eqref{equilibrium-distribution}
for the equilibrium distribution
\begin{equation}
f_a(x,p) = e^{\alpha_a - \theta\cdot p}
\label{f_a-eq}
\end{equation}
with arbitrary species-dependent chemical potentials $\alpha_a$
satisfies Eq.\ \eqref{Boltzmann-eq-mult},
as the collision term \eqref{Coll-psi-mult} vanishes.
Note that the four-vector $\theta^\mu$ is common to all species.

The equilibrium expressions \eqref{J_formula} and \eqref{T_formula}
for $J^\mu$ and $T^{\mu\nu}$ continue to apply,
with the species-dependent values of the Lorentz-violating coefficients.
This is also the case for the expressions \eqref{bar-theta} for $\bar{\theta}$,
\eqref{n} for $n$ and \eqref{u-mu} for $u^\mu$.
The four-vectors $u^\mu$ are proportional to $\theta^\mu$ for any species,
with species-dependent normalization $\bar{\theta}^{-1}$.
The temperature,
which is defined in the rest frame as $1/(k\theta^0)$,
is common to all particle species.

The total energy density and the pressure are given by
\begin{equation}
\mathcal{E} = \sum_{a=1}^n \mathcal{E}_a\>,
\qquad
P = \sum_{a=1}^n P_a
\end{equation}
where $\mathcal{E}_a$ and $P_a$ correspond to the partial energy density and
the partial pressure of the species $a$ given by formulas
\eqref{E_MB} and \eqref{P-n-theta_MB}.

Next consider the case where there are also inelastic collisions,
corresponding to chemical or nuclear reactions.
Following \cite{CercignaniKremer},
we will consider the case of four species $a = A, B, C, D$
subject to the reaction
\begin{equation}
A + B \> \rightleftarrows \> C + D \>.
\label{reaction_ABCD}
\end{equation}
The right-hand side of Eq.\ \eqref{Boltzmann-eq-mult} then includes the
additional term $R_a$ defined as
\begin{align}
R_A &= \frac{g}{h^3}\int \frac{d^3p_B}{\tilde{p}_{0B}}\frac{d^3p'_C}{\tilde{p}'_{0C}}\frac{d^3p'_D}{\tilde{p}'_{0D}}
(f_{C'}f_{D'} - f_A f_B) W(p_A,p_B\rightarrow p_C',p_D') \\
R_C &= \frac{g}{h^3}\int \frac{d^3p_A}{\tilde{p}_{0A}}\frac{d^3p_B}{\tilde{p}_{0B}}\frac{d^3p'_D}{\tilde{p}'_{0D}}
(f_A f_B - f_{C'}f_{D'}) W(p_C',p_D'\rightarrow p_A,p_B)
\end{align}
and analogous expressions for $R_B$ and $R_D$.
Similarly, the collisional functional \eqref{Coll-psi-mult} 
gets an additional term	
\begin{equation}
\int \psi_a R_a \frac{d^3p_a}{\tilde{p}_{0a}}\>.
\end{equation}
It then follows that the $J^\mu_a$ are no longer separately conserved,
but satisfy
\begin{equation}
\partial_\mu J_a^\mu = \nu_a l
\end{equation}
where
\begin{align}
l &= \int R_A(p_A) \frac{d^3p_A}{\tilde{p}_{0A}}
= - \int R_C(p'_C) \frac{d^3p'_C}{\tilde{p}'_{0C}} \nonumber\\
&= \frac{g}{h^3}\int \frac{d^3p_A}{\tilde{p}_{0A}}\frac{d^3p_B}{\tilde{p}_{0B}}\frac{d^3p'_C}{\tilde{p}'_{0C}}\frac{d^3p'_D}{\tilde{p}'_{0D}}
(f_{C'}f_{D'} - f_A f_B) W(p_A,p_B\rightarrow p_C',p_D')
\end{align}
and
\begin{equation}
\nu_A = \nu_B = -\nu_C = -\nu_D = 1\>.
\end{equation}
We see that the total current $J^\mu = \sum_{a=A}^D J_a^\mu$ is conserved.
This is also the case for the total energy-momentum tensor
$T^{\mu\nu} = \sum_{a=A}^D T_a^{\mu\nu}$, as can be shown by summing the
transfer equation over all species and substituting $\psi_a = p_a^\mu$.

However, if we substitute the ansatz \eqref{f_a-eq} into the (modified) transfer
equation, the extra term in the collisional functional no longer vanishes,
due to the fact that
\begin{equation}
f_{C'}f_{D'} - f_A f_B = \left(e^{\alpha_C + \alpha_D} - e^{\alpha_A + \alpha_A}\right)
e^{-\theta\cdot(p'_{C} + p'_{D})}
\end{equation}
Evidently, for this to vanish one needs the equilibrium condition
\begin{equation}
\alpha_A + \alpha_B - \alpha_C - \alpha_D = \sum_{a_A}^D \nu_a \alpha_a = 0
\end{equation}
relating the chemical potentials of the 4 particle species.

\section{Bose-Einstein condensation}
\label{sec:Bose-Einstein-condensation}

As an application we investigate the effects of the Lorentz-violating coefficients
on Bose-Einstein condensation for a monospecies gas.
Relativistic Bose-Einstein condensation without Lorentz violation has been worked
out by Landsberg and Dunning-Davies \cite{LandsbergDunning-Davis}
(see also \cite{CercignaniKremer}).
For simplicity we will consider $a^\mu = 0$, with nonzero $c^{\mu\nu}$.

As a starting point we take the expression for the particle density for the case $\epsilon = 1$,
which equals the zero-component of the current density given by Eq.\ \eqref{J_formula}:
\begin{equation}
J^0 = \frac{4\pi g m^3 u^0}{h^3 |\,\delta + c\,|} J_{21}(m\bar{\theta},\alpha) \>.
\label{particle-density-BE}
\end{equation}
where we used definition \eqref{u-mu}.

\begin{figure}[h]
 \begin{minipage}[c]{0.6\textwidth}
    \includegraphics[width=7cm]{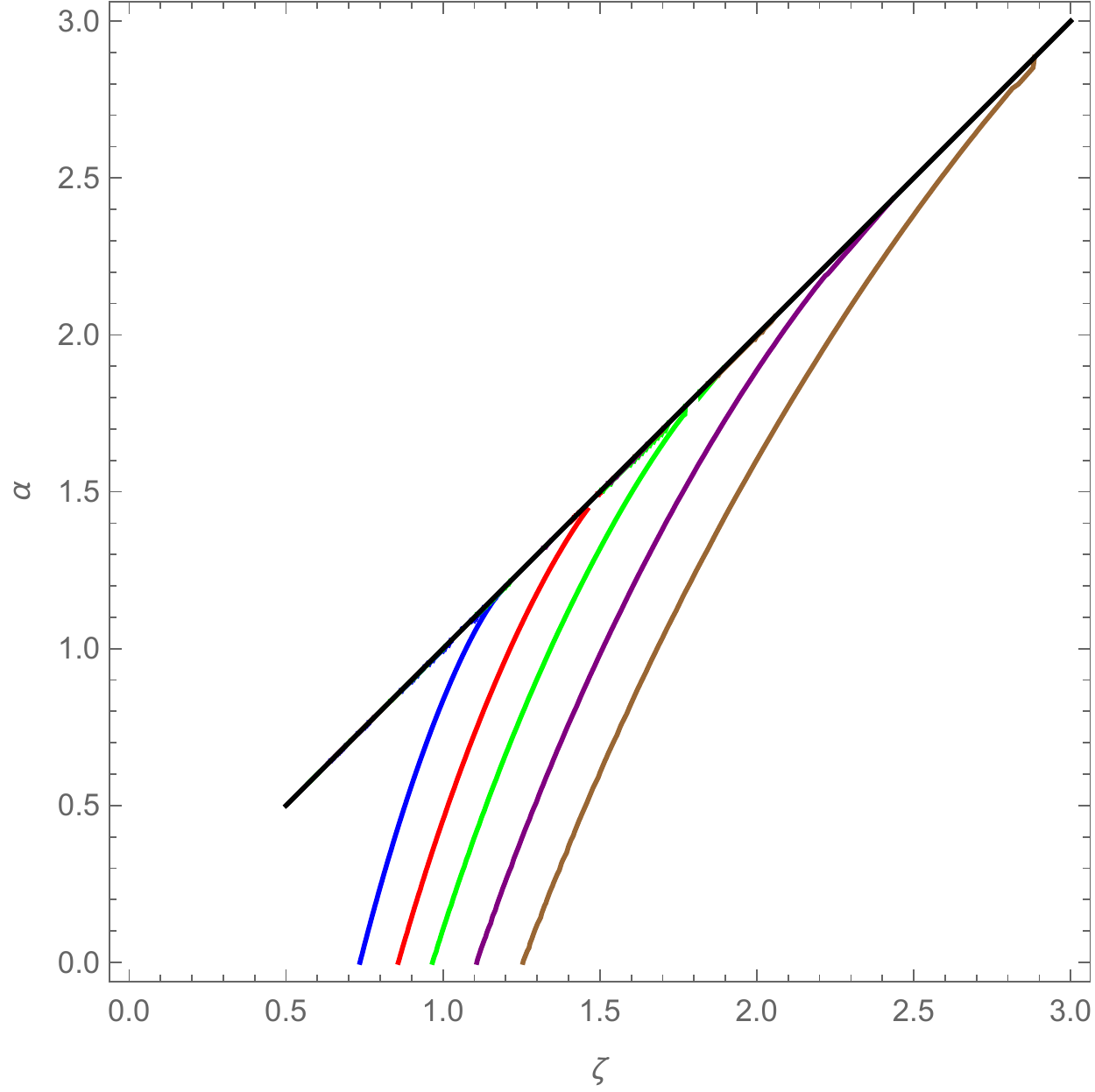}
  \end{minipage}\hfill
  \begin{minipage}[c]{0.4\textwidth}
   \caption{\small
Curves in the  $(\zeta,\alpha)$ plane
on which $J_{21}(\zeta,\alpha)$ equals, from left to right,
5, 3, 2, 1.25 and 0.8,
each one ending on the the line $\alpha = \zeta$.}
\label{fig:J21}
  \end{minipage}
\end{figure}

Denoting $m\bar{\theta} = \zeta$ we have
\begin{align}
J_{21}(\zeta,\alpha) &= \int_0^\infty \frac{\sinh^2x \cosh x}{e^{-\alpha + \zeta\cosh x} - 1}dx\nonumber\\
&= \frac{1}{\zeta^3}\int_0^\infty \frac{\sqrt{u^2 + 2\zeta u}\,(u + \zeta)}{e^{-\alpha + \zeta + u} - 1}du \nonumber\\
&= \int_0^\infty \frac{\sqrt{y^2 + 2y}\,(y + 1)}{e^{-\alpha + \zeta(y + 1)} - 1}dy
\label{J21-reps}
\end{align}
where we introduced the variables $u$ and $y$ defined by
\begin{equation}
y = \frac{u}{\zeta} = \cosh x - 1\>.
\end{equation}
Note first of all that $J_{21}(\zeta,\alpha)$ is only well defined for $\zeta \ge \alpha$.
Now let us fix the particle density $J^0$ to some value.
Working in the 
rest frame of the fluid,
Eq.\ \eqref{particle-density-BE} represents an implicit relation between the
variables $\zeta = m\bar{\theta} = m/(u^0 kT)$ and $\alpha = \mu_E/(kT)$.
From the third identity of Eq.\ \eqref{J21-reps} it follows that:
\begin{itemize}
\item For fixed $\alpha$, $J_{21}(\zeta,\alpha)$ is monotonically decreasing as a function of $\zeta$;
\item for fixed $\zeta$, $J_{21}(\zeta,\alpha)$ is monotonically increasing as a function of $\alpha$;
\item for $\zeta > 0$, $J_{21}(\zeta,\zeta)$ is finite and monotonically decreasing as a function of $\zeta$,
tending to zero if $\zeta \to \infty$ and tending to $\infty$ if $\zeta \to 0$.
\end{itemize}
This means that in the $(\zeta,\alpha)$ plane the tangents to the curves
defined by fixed values of $J^0$ always have a slope that is strictly larger than unity.
Therefore these curves will intersect the $45^\circ$ line $\alpha = \zeta$ at some point
and cease there, defining a critical, maximum value $\zeta_c$ for $\zeta$,
corresponding to a minimum, critical value $T_c$ for the temperature.
Below this temperature relation \eqref{particle-density-BE} between the particle density
and the temperature cannot be maintained.
In Fig.\ \ref{fig:J21} we have plotted, for illustration, a number of curves in the $(\zeta,\alpha)$ plane
on which $J_{21}(\zeta,\alpha)$ is constant,
each one ending on the the line $\alpha = \zeta$.

The physical interpretation of this effect is of course well known:
below $T_c$ a finite number of particles condenses in the lowest-energy state.
Denoting the total number of particles at the temperature $T$ as $N = VJ^0$ 
(with $V$ equal to the volume) and $N_0$ as the number of particles in the lowest-energy state,
we have for $T < T_c$ from Eq.\ \eqref{particle-density-BE}
\begin{equation}
N = N_0 + \frac{4\pi V g m^3 u^0}{h^3 |\,\delta + c\,|} J_{21}(\zeta,\zeta)
= \frac{4\pi V g m^3 u^0}{h^3 |\,\delta + c\,|} J_{21}(\zeta_c,\zeta_c)
\end{equation}
which yields the temperature dependence
\begin{equation}
\frac{N_0}{N} = 1 - \frac{J_{21}(\zeta,\zeta)}{J_{21}(\zeta_c,\zeta_c)}\>.
\label{critical-behavior}
\end{equation}
%
We can obtain explicit expressions for $T_c$ and $N_0/N$ in two limits:
\begin{enumerate}
\item In the nonrelativistic regime $\zeta \gg 1$ it follows from the second identity of
Eq.\ \eqref{J21-reps} that
\begin{equation}
J_{21}(\zeta,\zeta) \approx \sqrt{\frac{2}{\zeta^3}}\int_0^\infty \frac{\sqrt{u}\,du}{e^u - 1}
= \sqrt{\frac{\pi}{2\zeta^3}}\,\zeta_R(3/2)
\label{J21_zetazeta}
\end{equation}
where $\zeta_R(s)$ is the Riemann $\zeta$ function.
The critical temperature can now be obtained by inverting relation \eqref{particle-density-BE}.
One finds
\begin{align}
T_c &= \frac{h^2}{2\pi mk (u^0)^{5/3}}\left(\frac{J^0 |\,\delta+c\,|}{\zeta_R(3/2)}\right)^{2/3}\nonumber\\
&= \frac{h^2}{2\pi m^* k}\left(\frac{J^0 }{\zeta_R(3/2)}\right)^{2/3}
\left(\frac{|\,\delta+c\,|}{u^0}\right)^{2/3}
\label{T_c}
\end{align}
where in the second line we substituted the effective rest mass \eqref{effective-mass}.
The last factor in the second equation of \eqref{T_c} represents the correction due to
the coefficients $c^{\mu\nu}$; when evaluated to first order, it yields%
\footnote{The correction \eqref{T_c-correction-first-order} is consistent
with the corresponding one found in Ref.\ \cite{ColladayMcDonald}.}
\begin{equation}
\left(\frac{|\,\delta+c\,|}{u^0}\right)^{2/3} \approx 1 - \tfrac{2}{3}\text{Tr}[c^{ij}]\>.
\label{T_c-correction-first-order}
\end{equation}
For the temperature dependence \eqref{critical-behavior} we obtain from Eq.\ \eqref{J21_zetazeta}
in the nonrelativistic limit the well-known result
\begin{equation}
\frac{N_0}{N} = 1 - \left(\frac{T}{T_c}\right)^{3/2}\>.
\end{equation}
\item In the ultra-relativistic limit $\zeta \ll 1$ we obtain from the second identity of
Eq.\ \eqref{J21-reps}
\begin{equation}
J_{21}(\zeta,\zeta) \approx \frac{1}{\zeta^3}\int_0^\infty \frac{u^2\,du}{e^u - 1}
= \frac{2}{\zeta^3}\,\zeta_R(3)\>.
\label{J21_zetazeta_relativistic}
\end{equation}
Inverting relation \eqref{particle-density-BE} now yields
\begin{equation}
T_c = \frac{h}{2k}\left(\frac{J^0 }{\pi g\zeta_R(3)}\right)^{1/3}
|\,\delta+c\,|^{1/3}(u^0)^{-4/3}
\end{equation}
while the temperature dependence
\eqref{critical-behavior} reduces to the usual result
\begin{equation}
\frac{N_0}{N} = 1 - \left(\frac{T}{T_c}\right)^3\>.
\end{equation}
\end{enumerate}

It is interesting to identify the nature of the condensed state.
In the Lorentz-invariant case, it corresponds,
in the 
rest frame of the fluid,
to the state with zero three-momentum and zero velocity.
In the Lorentz-violating case the situation requires more careful consideration.
From Eq.\ \eqref{J21-reps}, or equivalently Eq.\ \eqref{I-reps},
we see that the condensed state will be the one in which $\theta_\mu p^\mu$ takes its minimum value.
Moreover, we have to take into account that $p^\mu$ is constrained to satisfy
the dispersion relation Eq.\ \eqref{dispersion-relation}.
We can solve this in the standard way by introducing a Lagrange multiplier
$\lambda$, define the function
\begin{equation}
H_c(p^\mu,\lambda) = \theta_\mu p^\mu + \lambda D(p^\mu)
\end{equation}
and solve the system
\begin{equation}
\frac{\partial H_c}{\partial p^\mu} = \frac{\partial H_c}{\partial \lambda} = 0\>. 
\end{equation}
Noting that $\partial D(p^\mu)/\partial p_\mu = 2\tilde{p}^\mu$
(with $\tilde{p}_\mu$ defined by Eq.\ \eqref{tilde-p}),
it follows that $\tilde{p}^\mu \propto \theta^\mu$.
Using definitions \eqref{u-mu} and \eqref{u-normalization} for $u^\mu$,
together with the constraint \eqref{constraint},
we readily conclude that the four-velocity $\dot{x}^\mu$ of the condensed state
is exactly equal to $u^\mu$.
Therefore, the condensate has zero three-velocity in the rest frame of the fluid.
Note that its spatial momentum is not zero in that frame, 
but is defined by Eq.\ \eqref{pmu_dot-x}.

\section{Conclusions}
\label{sec:conclusions}

In this work we presented a treatment of relativistic kinetic theory
in the presence of Lorentz-violating dispersion relations associated
to two commonly considered sets of coefficients in the SME.
Adopting an manifestly observer-covariant approach,
we carefully defined the phase-space distribution function
and the Boltzmann equation, for the cases of classical and quantum (fermionic or bosonic)
statistics.
Using the associated transfer equation,
we generalized Boltzmann's H-theorem to the Lorentz-violating case,
using an appropriate definition of the entropy current.

Next we analyzed the equilibrium solutions of the Boltzmann equation.
Concentrating on those in which the phase-space distribution function is independent
of the spacetime coordinates,
we first constructed the appropriate forms of the particle-density current
and the energy-momentum tensor.
Using these, we then extracted the standard thermodynamic variables
such as the energy density, the isotropic pressure, the temperature,
the chemical potential and the entropy.
We found that these quantities are defined in a fashion that
is parallel to the Lorentz-symmetric case,
but with crucial factors accounting for Lorentz-violating effects associated with
the SME parameters.
As an application, we analyzed Bose-Einstein condensation in the presence
of Lorentz violation.
We also generalized the Boltzmann equation to the case of various species
of particles with different sets of Lorentz-violating coefficients,
allowing for the possibility of a chemical reaction in nonelastic collisions.
We compared our results to those obtained previously in the literature
in studies using a conventional approach to thermodynamics
in the presence of Lorentz violation, based on a partition function.

While we applied our approach explicitly on to the sets of SME coefficients
$a^\mu$ and $c^{\mu\nu}$, our general setup can be straightforwardly adapted
to accommodate dispersion relations associated to any of the other coefficients
in the minimal or nonminimal SME,
as well as other models with SME-like Lorentz-violating dispersion relations.
Given any dispersion relation $D(p) = 0$ involving Lorentz-violating coefficients,
as an alternative to Eq.\ \eqref{dispersion-relation},
one defines
\begin{equation}
\tilde{p}_\mu = \frac{\partial D}{\partial p^\mu}\>.
\label{new-tilde-p-mu}
\end{equation}
The modified Boltzmann equation and the Uehling-Uhlenbeck equation are then defined
by Eqs.\ \eqref{Boltzmann-eq} and \eqref{UU-eq},
with the new definition \eqref{new-tilde-p-mu} of $\tilde{p}^\mu$.
This also holds for the subsequent analysis.
What needs to be adapted on a case-by-case basis is the
calculation of the thermodynamical quantities in section \ref{sec:particle-current_EM-tensor}.
It may not be feasible to compute the relevant integral \eqref{generating-function}
for all cases.
If not, it may be necessary instead to resort to a simplified analysis
to first order in the Lorentz-violating coefficients.

In this work we have presented the basic framework for
kinetic theory in the presence of Lorentz-violating dynamics,
and an analysis of equilibrium solutions of the Boltzmann equation.
There are various relevant issues that are beyond the scope
of the current work.
The most obvious one is the application of the Boltzmann equation
to analyze out-of-equilibrium processes in the presence
of Lorentz violation.
In particular, it should be interesting to develop the Chapman-Enskog method
to derive the Lorentz-violating effects on the equations of hydrodynamics,
such as the laws of Navier-Stokes and Fourier.
Another interesting open topic is the derivation and analysis
of the Boltzmann equation with Lorentz violation
in the presence of gravitational fields \cite{Kostelecky03}.

\acknowledgments
It is a pleasure to thank Jorge Diaz and Alan Kosteleck\'y for discussions
and helpful suggestions.
Financial support from Funda\c c\~ao para a Ci\^encia e a Tecnologia -- FCT
through the project No.\ UIDB/00099/2020 is gratefully acknowledged.

\end{document}